\documentclass[sigconf]{acmart}

\copyrightyear{2026}
\acmYear{2026}
\setcopyright{cc}
\setcctype{by}
\acmConference
    [CIKM '26]
    {Proceedings of the 35th ACM International Conference on Information and Knowledge Management}
    {November 07--11, 2026}
    {Rome, Italy}
\acmBooktitle
    {Proceedings of the 35th ACM International Conference on Information and Knowledge Management (CIKM '26), November 07--11, 2026, Rome, Italy}
\acmDOI{10.1145/3799682.3841117}
\acmISBN{979-8-4007-2539-5/2026/11}

\usepackage{enumitem}
\usepackage{amsmath}
\usepackage{multirow}
\usepackage{subcaption}

\begin{document}

\title
    [GOD: Enhancing Generalization via Deep Grafting for Sequential Recommendation]
    {GOD: Enhancing Generalization via Deep Grafting\\for Sequential Recommendation}

\author{WooJoo Kim}
\email{kimuj0103@postech.ac.kr}
\affiliation
{
    \institution{Pohang University of\\Science and Technology}
    \city{Pohang}
    \country{Republic of Korea}
}
\author{JunYoung Kim}
\email{junyoungkim@postech.ac.kr}
\affiliation
{
    \institution{Pohang University of\\Science and Technology}
    \city{Pohang}
    \country{Republic of Korea}
}
\author{JaeHyung Lim}
\email{jaehyunglim@postech.ac.kr}
\affiliation
{
    \institution{Pohang University of\\Science and Technology}
    \city{Pohang}
    \country{Republic of Korea}
}
\author{HwanJo Yu}
\authornote{Corresponding author.}
\email{hwanjoyu@postech.ac.kr}
\affiliation
{
    \institution{Pohang University of\\Science and Technology}
    \city{Pohang}
    \country{Republic of Korea}
}

\begin{abstract}
    Sequential recommenders often struggle with sparse and noisy histories, limiting generalization to unseen interactions.
    Knowledge distillation mitigates this by transferring dense supervision from a teacher to a student.
    However, most distillation methods run teacher and student independently, then match student outputs or representations to the teacher.
    Such supervision entangles student-component effects, blurring whether weak generalization stems from unreliable embeddings, overfitted encoding, or co-adaptation to sparse histories.
    In this paper, we propose \textbf{G}raft-\textbf{O}riented \textbf{D}istillation (\textbf{GOD}), a component-level distillation framework for improved generalization through grafting.
    Grafting denotes replacing selected frozen-teacher components with trainable student counterparts to build hybrid source models.
    GOD uses these hybrid models to evaluate student embeddings with the teacher encoder and the student encoder with teacher embeddings, providing component-level feedback.
    At inference, GOD uses only the student, incurring no additional cost.
    Across three real-world datasets, GOD outperforms state-of-the-art baselines by up to 13.92\%.
\end{abstract}



\maketitle

\section{Introduction}

Sequential recommendation (SR) aims to predict the next item a user will interact with by modeling sequential patterns in interaction histories \cite{wang2015learning, wu2019session, lee2026capturing, kim2026overlooked}.
Deep sequential models \cite{hidasi2015session, li2017neural, tan2016improved, liu2018stamp, tuan20173d}, particularly Transformer-based recommenders \cite{kang2018self, sun2019bert4rec, ying2018sequential, li2020time, fan2022sequential, liu2023linrec}, have achieved strong performance by capturing complex dependencies among past interactions.
However, real-world user histories are often sparse and noisy \cite{zhou2020s3, wang2021denoising, liu2021contrastive}, making SR models prone to fitting observed co-occurrences or spurious transitions rather than reliable patterns for future interactions \cite{yao2021self, liu2021augmenting}.
Therefore, a central challenge in SR is learning robust sequential representations that generalize beyond limited and noisy training histories \cite{xie2022contrastive, qiu2022contrastive, dang2023uniform, du2023ensemble}.

Knowledge distillation (KD) \cite{hinton2015distilling, romero2014fitnets} can alleviate data sparsity by transferring dense supervision from a pretrained teacher to a compact student.
Existing recommendation KD methods have explored diverse forms of supervision, including ranking outputs \cite{tang2018ranking, lee2019collaborative, kang2020rrd}, embeddings \cite{chen2021scene, kang2021topology}, interest-level knowledge \cite{du2024multi}, and self-distilled signals \cite{yuan2021improving, du2023ensemble, wei2024leave, lim2025federated}.
Despite their effectiveness, most methods use the teacher primarily as an external signal generator.
The teacher first produces predictions or representations through its own inference path, and the student is then trained to match these signals through a separate inference path.

KD with separated inference paths is particularly limiting in SR since sparsity and noise affect the components that form sequence representations.
Most SR models map items into embeddings and transform the resulting sequence with an encoder \cite{hidasi2015session, zhou2022filter, kang2018self}.
Sparse and noisy interactions can yield unreliable student embeddings, while the encoder can amplify spurious dependencies in short histories.
In a student-only path, embedding and encoder errors can be entangled, making it difficult to provide component-specific guidance for generalization.
Thus, KD can better support generalization through component-level feedback beyond matching teacher-produced signals from separately executed paths.

\begin{figure}[t]
    \centering
    \includegraphics[width=\linewidth]{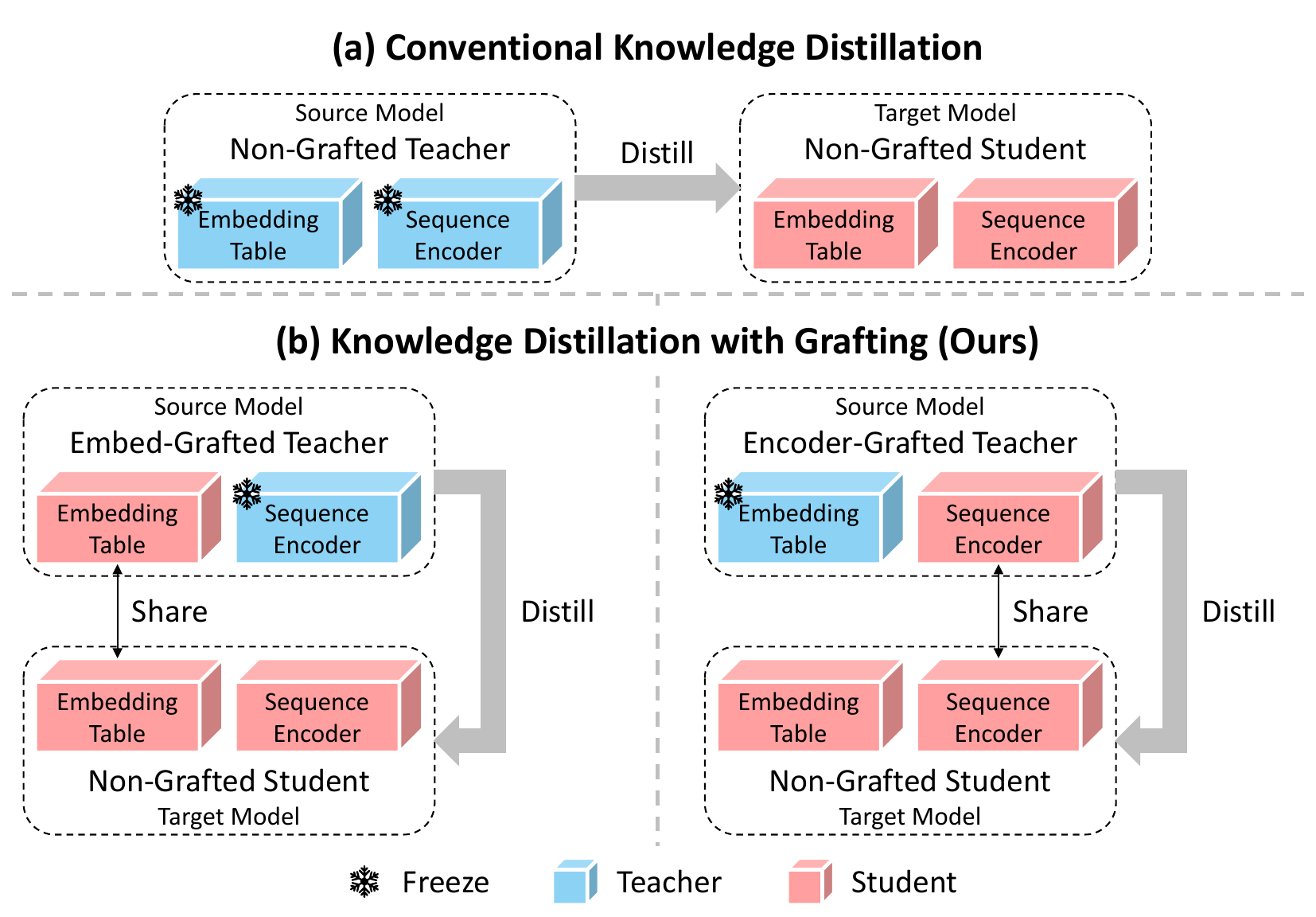}
    \caption{Illustration of (a) conventional KD with an independent teacher and (b) KD with grafting. In (b), hybrid source models replace selected teacher components with student counterparts. Inference uses only the non-grafted student.}
    \label{fig/graft}
    \Description{}
\end{figure}

Intermediate-representation KD \cite{romero2014fitnets, jiao2020tinybert} can partially mitigate this issue, yet it still compares targets from separate teacher and student paths.
Such feedback blurs whether weak generalization stems from unreliable embeddings, limited encoding, or their co-adaptation to sparse histories.
A more direct strategy is to replace a single teacher component with its student counterpart while keeping the rest teacher-side.
The resulting hybrid path tests student embeddings under the teacher encoder and the student encoder under teacher embeddings, yielding component-level feedback for enhanced generalization.
This aligns with model \textit{grafting}, which combines components from different networks within a single computation path \cite{meng2020filter, shen2021progressive, park2024grafting}.
As illustrated in \autoref{fig/graft}, we use grafting in KD to isolate and supervise student components during training.

Based on grafting, we propose \textbf{G}raft-\textbf{O}riented \textbf{D}istillation (\textbf{GOD}), a component-level KD framework for SR.
GOD constructs hybrid source models by grafting the student embedding table or the student sequence encoder into a frozen teacher, while the non-grafted student serves as the target model.
These grafted sources expose different student components to teacher-side computation, providing complementary supervision for generalization.
For Transformer-based SR models, \textit{Grafted Encoding} stabilizes representation generation by enabling mutual attention between teacher- and student-side tokens.
GOD then distills the representations generated by the source models with \textit{Graft-aware Contrastive Learning}, which adaptively balances correlated grafted views.
During inference, GOD uses only the non-grafted student, adding no extra inference cost.

Our main contributions are summarized as follows:
\begin{itemize}[itemsep=1pt, left=0pt]
    \item We revisit KD for SR from a component-level perspective, highlighting how separated teacher-student paths can weaken generalization under sparse histories.
    \item We propose GOD, a component-level KD framework that constructs hybrid source models by grafting trainable student components into a frozen teacher for enhanced generalization.
    \item Experiments on three real-world datasets demonstrate that GOD consistently outperforms existing KD and self-supervised baselines by up to 13.92\%.
\end{itemize}

\section{Preliminary}

\subsection{Problem Formulation}

In SR, the goal is to predict the next item from the historical interaction sequence of a user.
Let $\mathcal{U}$ and $\mathcal{I}$ denote the user and item sets, respectively.
Each user $u \in \mathcal{U}$ has a chronologically ordered sequence $s_u=[i_{u, 1}, i_{u, 2}, \ldots, i_{u, |s_u|}]$, where $i_{u, t} \in \mathcal{I}$ is the $t$-th item interacted with by user $u$.
Given $s_u$, the prediction is formulated as:
\begin{equation}
    \underset{i \in \mathcal{I}}{\operatorname{argmax}} \ p(i_{u, |s_u|+1} = i | s_u).
\end{equation}

\subsection{Sequential Recommender}

Most sequential recommenders \cite{hidasi2015session, tang2018personalized, kang2018self, zhou2022filter} consist of two components: an embedding table and a sequence encoder.
Among them, Transformer-based models \cite{zhou2020s3, liu2021contrastive, chen2022intent, qin2024intent, wang2024relative, zhang2025frequency} have become particularly prevalent due to their effectiveness in modeling complex sequential patterns \cite{vaswani2017attention, devlin2018bert}.
We therefore describe the standard Transformer-based formulation used in SR.

\subsubsection{Embedding Table}
Two learnable embedding tables are used: an item table $I \in \mathbb{R}^{|\mathcal{I}| \times d}$ and a position table $P \in \mathbb{R}^{N \times d}$, where $N$ is the maximum sequence length.
After truncating the earliest items or front-padding zeros to length $N$, $s_u$ is embedded as:
\begin{equation}
    e_u = [I_{i_{u, 1}} + P_1, I_{i_{u, 2}} + P_2, \ldots, I_{i_{u, N}} + P_N] \in \mathbb{R}^{N \times d},
\end{equation}
where $I_{i_{u, t}}$ and $P_t$ are the embeddings of item $i_{u, t}$ and position $t$.

\subsubsection{Sequence Encoder}
The sequence embedding $e_u$ is processed by an $L$-layer multi-head Transformer encoder, where $\mathrm{Trm}^l(\cdot)$ denotes the $l$-th layer.
With $H^0_u = e_u$, each layer outputs:
\begin{equation}
    H^l_u = [h^l_{u, 1}, h^l_{u, 2}, \ldots, h^l_{u, N}] = \mathrm{Trm}^l(H^{l-1}_u) \in \mathbb{R}^{N \times d}.
\end{equation}
The final-layer last-position representation $h^L_{u, N}$ defines the sequence representation $h_u = f(e_u)$, where $f(\cdot)$ applies all Transformer layers and selects the last position.

\subsection{Next Item Prediction}
Given $s_u$ and ground-truth item index $g_u$, the next-item probability distribution and recommendation loss are computed as:
\begin{equation}
    \begin{aligned}
        \hat{y}_u & = \mathrm{Softmax}(h_u I^\top) \in \mathbb{R}^{|\mathcal{I}|}, \\
        \mathcal{L}_{rec}(u) & = -\log{\hat{y}_u[g_u]}.
    \end{aligned}
\end{equation}

\section{Grafting Analysis for Distillation} \label{sec/grafting_analysis_for_distillation}
\begin{figure}[t]
  \centering
  \begin{subfigure}[t]{0.23\textwidth}
    \centering
    \includegraphics[width=\textwidth]{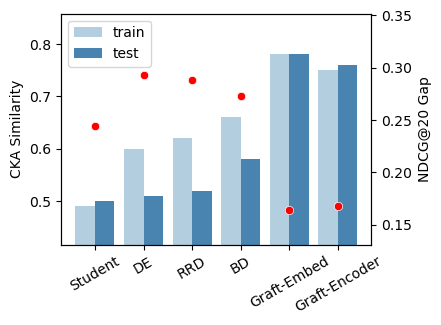}
    \caption{Amazon Beauty}
  \end{subfigure}
  \hfill
  \begin{subfigure}[t]{0.23\textwidth}
    \centering
    \includegraphics[width=\textwidth]{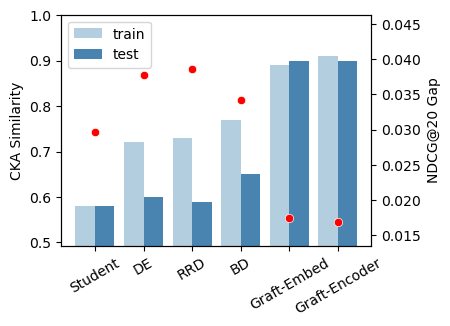}
    \caption{MovieLens 1M}
  \end{subfigure}
  \caption{Train/test CKA similarity of teacher-student sequence representations and NDCG@20 gaps with SASRec.}
  \label{fig/graft_generalization}
  \Description{}
\end{figure}

\begin{figure*}[t]
  \centering
  \begin{subfigure}[t]{\textwidth}
    \centering
    \includegraphics[width=0.18\textwidth]{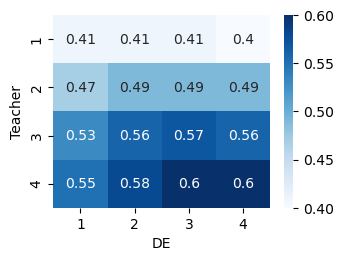}
    \includegraphics[width=0.18\textwidth]{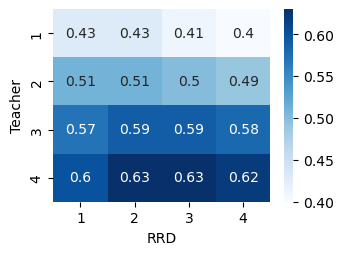}
    \includegraphics[width=0.18\textwidth]{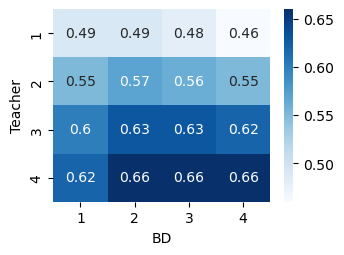}
    \includegraphics[width=0.18\textwidth]{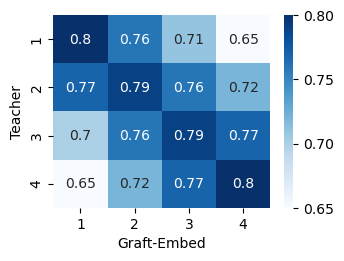}
    \includegraphics[width=0.18\textwidth]{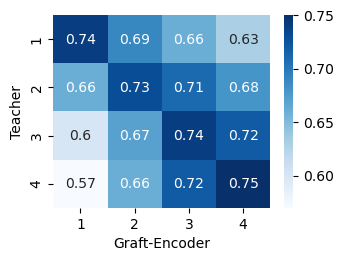}
    \caption{Amazon Beauty}
  \end{subfigure}
  \hfill
  \begin{subfigure}[t]{\textwidth}
    \centering
    \includegraphics[width=0.18\textwidth]{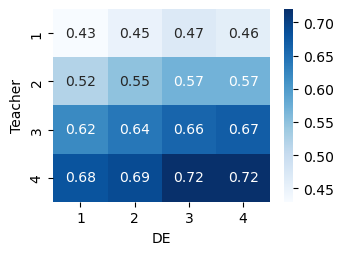}
    \includegraphics[width=0.18\textwidth]{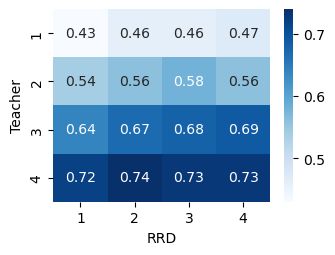}
    \includegraphics[width=0.18\textwidth]{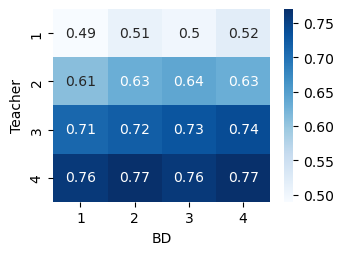}
    \includegraphics[width=0.18\textwidth]{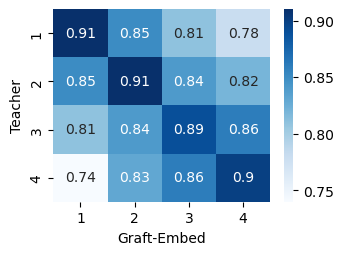}
    \includegraphics[width=0.18\textwidth]{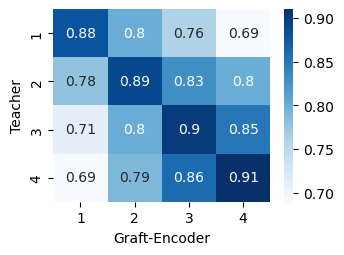}
    \caption{MovieLens 1M}
  \end{subfigure}
  \caption{Layer-wise CKA heatmaps of teacher-student intermediate representations with 4-layer SASRec.}
  \label{fig/graft_alignment}
  \Description{}
\end{figure*}

Before presenting GOD, we analyze whether grafting can provide component-level feedback for generalization.
We evaluate whether grafting improves the transfer of distilled knowledge to unseen sequences and encourages procedural alignment between teacher and student encoding processes.

\subsection{Generalization of Distilled Knowledge}

Grafting improves generalization through component-level feedback.
Conventional KD supervises the full student path with signals from an independent teacher, jointly optimizing student embeddings and encoder.
Grafting instead replaces a frozen teacher component with its trainable student counterpart, forming a hybrid source model that evaluates the student component under teacher-side computation.
Since the grafted component is shared with the student, KD regularization on the student also refines the source model \cite{hinton2015distilling, saglietti2022solvable}, creating an adaptive supervision loop.
This resembles bi-directional KD \cite{zhang2018deep, kweon2021bidirectional} but requires no explicit teacher updates and provides finer-grained guidance than output-level imitation.

\noindent
\textbf{Empirical Evidence.}
We compare Student (no KD) and five KD variants (DE \cite{kang2020rrd}, RRD \cite{kang2020rrd}, BD \cite{kweon2021bidirectional}, Graft-Embed, and Graft-Encoder\footnote{Graft-Embed and Graft-Encoder use the \textit{Embed-Grafted Teacher} and \textit{Encoder-Grafted Teacher} in \autoref{fig/graft} as their source models, respectively.}) using SASRec \cite{kang2018self} as the backbone.
\autoref{fig/graft_generalization} reports centered kernel alignment (CKA) \cite{kornblith2019similarity} between teacher and student representations on train/test sequences, along with NDCG@20 gaps (train minus test).
Student shows low CKA without KD, while DE and RRD improve train alignment but suffer test-alignment drops and large NDCG@20 gaps.
BD partially reduces these gaps through dynamic distillation, but Graft-Embed and Graft-Encoder maintain high train/test CKA with much smaller NDCG@20 gaps, indicating better generalization of distilled knowledge.

\subsection{Procedural Alignment}

Beyond generalization, we analyze procedural alignment, i.e., layer-wise correspondence between teacher and student encoding processes.
Conventional KD supervises the student with teacher signals from separate paths, which can encourage output imitation rather than aligned encoding behavior \cite{shen2021progressive}.
Distilling intermediate representations \cite{romero2014fitnets, jiao2020tinybert} or attention maps \cite{zhou2023attention} may expose internal teacher states, but these targets are still produced by separate teacher forward passes.
Grafting instead places trainable student components inside teacher-conditioned paths, guiding student encoding without explicitly matching every internal teacher state.

\noindent
\textbf{Empirical Evidence.}
\autoref{fig/graft_alignment} visualizes CKA heatmaps of teacher--student intermediate representations on training sequences with 4-layer SASRec.
DE, RRD, and BD show teacher-final-layer dominance, with all student layers most aligned to the final teacher layer, suggesting output imitation rather than layer-wise procedural alignment.
In contrast, Graft-Embed and Graft-Encoder show clear diagonal patterns, with student layers most aligned to corresponding teacher layers.
Notably, this correspondence emerges without explicit intermediate representation distillation, suggesting that grafting aligns student and teacher encoding processes through structural coupling rather than direct state matching.

\section{Proposed Framework: GOD}

\begin{figure}[t]
    \centering
    \includegraphics[width=\linewidth]{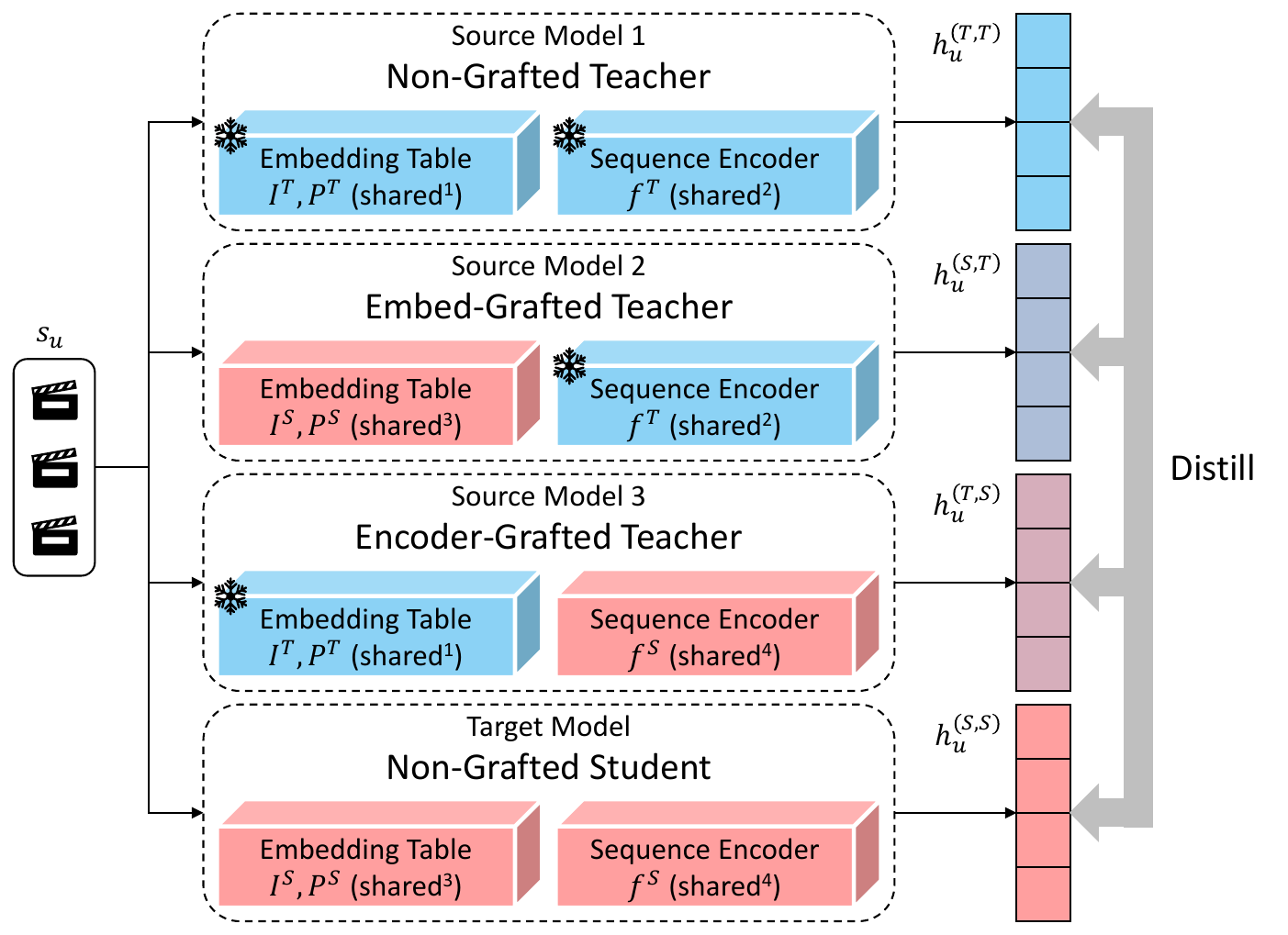}
    \caption{Illustration of GOD. GOD distills representations of three source models---\textit{Non-Grafted Teacher}, \textit{Embed-Grafted Teacher}, and \textit{Encoder-Grafted Teacher}---into \textit{Non-Grafted Student}. Identical ``shared'' indices indicate parameter sharing.}
    \label{fig/god}
    \Description{}
\end{figure}

We propose \textbf{GOD}, a KD framework using hybrid source models constructed via grafting, as illustrated in \autoref{fig/god}.
Each hybrid model replaces the teacher embedding table or sequence encoder with its student counterpart, exposing the student component to teacher-side computation.
GOD generates coupled representations with \textit{Grafted Encoding} and distills them with \textit{Graft-aware Contrastive Learning}.
At inference, GOD uses only the non-grafted student.

\subsection{Source / Target Model Configuration}

Following the grafting analysis in Sec.~\ref{sec/grafting_analysis_for_distillation}, GOD constructs three source models for component-level distillation: \textit{Non-Grafted Teacher}, \textit{Embed-Grafted Teacher}, and \textit{Encoder-Grafted Teacher}.
\textit{Non-Grafted Teacher} preserves the original teacher as a stable knowledge source.
\textit{Embed-Grafted Teacher} replaces the teacher embedding table with the student embedding table and feeds student embeddings to the teacher encoder.
\textit{Encoder-Grafted Teacher} replaces the teacher encoder with the student encoder and processes teacher embeddings.
The target model, \textit{Non-Grafted Student}, receives distilled sequence representations from all source models.
As shown in \autoref{fig/god}, all models share frozen-teacher and trainable-student components, introducing no standalone source-model parameters.

\noindent
\textbf{Forwarding Process.}
For a sequence $s_u$, we first construct teacher- and student-side embeddings.
Let $I^T, P^T$ and $I^S, P^S$ denote the item and position embedding tables of the teacher and student, respectively.
The embeddings are given by:
\begin{equation}
    \begin{aligned}
        e_u^T & = [I_{i_{u, 1}}^T + P_1^T, I_{i_{u, 2}}^T + P_2^T, \ldots, I_{i_{u, N}}^T + P_N^T] \in \mathbb{R}^{N \times d^T}, \\
        e_u^S & = [I_{i_{u, 1}}^S + P_1^S, I_{i_{u, 2}}^S + P_2^S, \ldots, I_{i_{u, N}}^S + P_N^S] \in \mathbb{R}^{N \times d^S}.
    \end{aligned}
\end{equation}
With $f^T(\cdot)$ and $f^S(\cdot)$ denoting the teacher and student sequence encoders, the two embeddings are paired with the two encoders to produce four representations:
\begin{equation}
    \begin{aligned}
        h_u^{(T,T)} & = f^T(e_u^T) \cdot W_{down} & \in \mathbb{R}^{d^S}, \\
        h_u^{(S,T)} & = f^T(e_u^S \cdot W_{up}) \cdot W_{down} & \in \mathbb{R}^{d^S}, \\
        h_u^{(T,S)} & = f^S(e_u^T \cdot W_{down}) & \in \mathbb{R}^{d^S}, \\
        h_u^{(S,S)} & = f^S(e_u^S) & \in \mathbb{R}^{d^S},
    \end{aligned}
\end{equation}
where $h^{(M_1,M_2)}_u$ denotes the representation generated using the embeddings of $M_1$ and the sequence encoder of $M_2$ for $M_1, M_2 \in \{T, S\}$.
Thus, $h_u^{(T,T)}$, $h_u^{(S,T)}$, $h_u^{(T,S)}$, and $h_u^{(S,S)}$ are generated by \textit{Non-Grafted Teacher}, \textit{Embed-Grafted Teacher}, \textit{Encoder-Grafted Teacher}, and \textit{Non-Grafted Student}, respectively.
GOD uses shared linear projections $W_{down} \in \mathbb{R}^{d^T \times d^S}$ and $W_{up} \in \mathbb{R}^{d^S \times d^T}$ to align dimensions while preserving the structural effect of grafting (see \autoref{fig/god_ablation_projection} in Sec.~\ref{sec/ablation_study}).

\subsection{Grafted Encoding}

\begin{figure}[t]
    \centering
    \includegraphics[width=\linewidth]{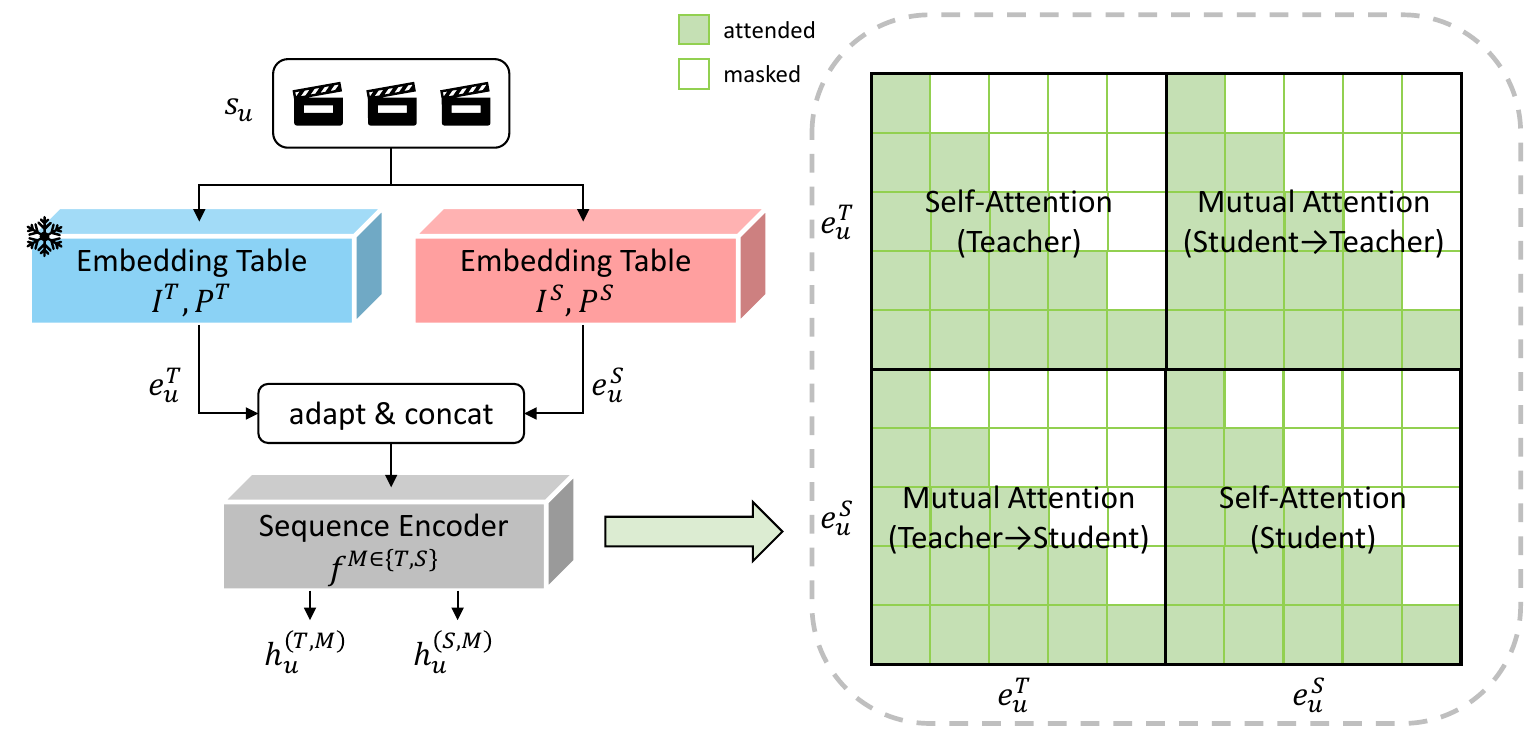}
    \caption{Illustration of Grafted Encoding with teacher-student mutual attention.}
    \label{fig/grafted_encoding}
    \Description{}
\end{figure}

GOD is applicable to SR models with separable embeddings and encoders, while Grafted Encoding (GE) targets Transformer-based models, which dominate recent SR research \cite{zhou2020s3, liu2021contrastive, chen2022intent, qin2024intent, wang2024relative, zhang2025frequency}.
GE stabilizes representation generation by concatenating teacher- and student-side embeddings and enabling mutual attention, as illustrated in \autoref{fig/grafted_encoding}.
Early in training, grafted student components are not yet well optimized, which can make hybrid source model representations unstable.
Mutual attention lets teacher-side embeddings provide stable context early on, while improved student-side embeddings later refine teacher-conditioned representations without updating the frozen teacher \cite{zhang2018deep, kweon2021bidirectional}.
Thus, GE strengthens structural coupling while keeping the teacher frozen.

\noindent
\textbf{Forwarding Process.}
In GE, the four models are grouped by their shared sequence encoder: \textit{Non-Grafted Teacher} and \textit{Embed-Grafted Teacher} share $f^T(\cdot)$, while \textit{Encoder-Grafted Teacher} and \textit{Non-Grafted Student} share $f^S(\cdot)$.
For each pair, teacher- and student-side embeddings are adapted to the same dimension, concatenated, and passed through the shared encoder.
Let $\mathrm{Trm}^{M,l}(\cdot)$ denote the $l$-th Transformer layer of encoder $f^{M \in \{ T, S \}}$, whose depth and hidden dimension are $L^M$ and $d^M$, respectively.
With $H^{T, 0}_u = [e^T_u; e^S_u \cdot W_{up}]$ and $H^{S, 0}_u = [e^T_u \cdot W_{down}; e^S_u]$, the forwarding process is:
\begin{equation}
    \begin{aligned}
        H^{M, l}_u & = [h^{M, l}_{u, 1}, h^{M, l}_{u, 2}, \ldots, h^{M, l}_{u, N}, h^{M, l}_{u, N+1}, \ldots, h^{M, l}_{u, 2N}] \\
        & = \text{Trm}^{M, l}(H^{M, l-1}_u) \in \mathbb{R}^{2N \times d^M}.
    \end{aligned}
\end{equation}
Instead of using separate causal attention for each side, GE enables mutual attention between teacher- and student-side tokens.
The causal attention mask is:
\begin{equation}
    (A_{causal})_{ij} = \mathbb{I}[j \le i], \quad A_{causal} \in \{0, 1\}^{N \times N},
\end{equation}
where $(A_{causal})_{ij} = 1$ allows the $i$-th query token to attend to the $j$-th key token.
For the concatenated sequence, GE uses:
\begin{equation}
    A_{GE} =
        \begin{bmatrix}
            A_{causal} & A_{causal} \\
            A_{causal} & A_{causal}
        \end{bmatrix}
    \in \{0, 1\}^{2N \times 2N}.
\end{equation}
Finally, the four sequence representations are obtained from the last valid positions of the teacher- and student-side token blocks:
\begin{equation}
    \begin{aligned}
        h_u^{(T,T)} & = h^{T, L^T}_{u, N} \cdot W_{down} & \in \mathbb{R}^{d^S}, \\
        h_u^{(S,T)} & = h^{T, L^T}_{u, 2N} \cdot W_{down} & \in \mathbb{R}^{d^S}, \\
        h_u^{(T,S)} & = h^{S, L^S}_{u, N} & \in \mathbb{R}^{d^S}, \\
        h_u^{(S,S)} & = h^{S, L^S}_{u, 2N} & \in \mathbb{R}^{d^S}.
    \end{aligned}
\end{equation}

\noindent
\textbf{Complexity.}
Although GE forms a length-$2N$ concatenated sequence, its overhead is limited to training-time attention since GE is not used during inference.
Without GE, four length-$N$ paths incur cost $O(2N^2d^T + 2N^2d^S)$, whereas GE uses two length-$2N$ paths with cost $O(4N^2d^T + 4N^2d^S)$, roughly doubling attention computation.
This overhead comes from stronger structural coupling and can be offset by faster convergence in practice (see \autoref{tab/efficiency} in Sec.~\ref{sec/efficiency_and_sensitivity}).
For memory cost, we also evaluate GE with each side sequence halved, so the concatenated sequence keeps length $N$.
GE still improves performance in this setting (see \autoref{tab/ablation_ge} in Sec.~\ref{sec/ablation_study}).

\subsection{Graft-aware Contrastive Learning}

Leveraging the four representations, GOD performs KD through contrastive learning.
Rather than point-wise matching or logit imitation, contrastive learning transfers relational structure among sequences across views \cite{oord2018representation, xie2022contrastive}.
This is well suited to GOD since representations of source and target models form structurally related views through shared teacher and student components.

GOD considers pairwise relations among all four representations \cite{qin2023meta, chong2023ct4rec, wang2024relative}.
We include both source-target and source-source pairs because hybrid source models contain trainable student components, allowing source-source relations to regularize the student embeddings or encoder.
With $\mathcal{A} = \{(T,T), (S,T), (T,S), (S,S)\}$, the contrastive loss for each pair is:
\begin{equation}
    \mathcal{L}_{pair}(h^a_u, h^b_u) = -\log \frac{\exp(h^a_u \cdot h^b_u / \tau)}{\sum_{s_{u'} \in B} \exp({h^a_u \cdot h^b_{u'} / \tau})},
\end{equation}
where $\tau$ is the temperature and $B$ is the mini-batch of sequences.

The six pair types, however, are not equally informative since the four representations share teacher and student components.
Following prior works on similarity-aware pair weighting \cite{wang2019multi, sun2020circle, kim2026flame}, GOD down-weights highly similar relations to reduce redundant supervision and emphasize complementary component-level signals.
Specifically, the pair weight and distillation objective are defined as:
\begin{equation}
    \begin{aligned}
        w(a, b) & = \frac{\exp(\mathrm{sg}(-\frac{1}{|B|} \sum_{s_u \in B} h^a_u \cdot h^b_u))}{\sum_{\{a', b'\} \subset \mathcal{A}} \exp({\mathrm{sg}(-\frac{1}{|B|} \sum_{s_u \in B} h^{a'}_u \cdot h^{b'}_u))}}, \\
        \mathcal{L}_{kd}(u) & = \sum_{\{a, b\} \subset \mathcal{A}} w(a, b) \mathcal{L}_{pair}(h^a_u, h^b_u),
    \end{aligned}
\end{equation}
where $\mathrm{sg}(\cdot)$ denotes stop-gradient.
GCL down-weights highly similar pair types and emphasizes less-aligned ones, while stop-gradient keeps $w(a,b)$ as an adaptive coefficient rather than a directly optimized target.

\subsection{Inference and Optimization}

\textbf{Inference.}
GOD uses only \textit{Non-Grafted Student} for inference:
\begin{equation}
    p(i_{u, |s_u|+1} | s_u) = \hat{y}_u = \mathrm{Softmax}(h_u {I^S}^\top),
\end{equation}
where $h_u = f^S(e_u^S)$.
Note that with GE, $h_u$ differs from $h^{(S,S)}_u$, which is generated with teacher-student mutual attention for training.

\noindent
\textbf{Optimization.}
Next-item prediction uses the sequence representation of \textit{Non-Grafted Student}, and KD improves its consistency and generalization.
Therefore, we jointly optimize the recommendation and distillation objectives with loss coefficient $\lambda$:
\begin{equation}
    \mathcal{L} = \sum_{u \in \mathcal{U}} ( \mathcal{L}_{rec}(u) + \lambda \mathcal{L}_{kd}(u) ).
\end{equation}

\section{Experiment}

In this section, we address the following research questions (RQs):
\begin{itemize}[itemsep=1pt, left=0pt]
    \item {\bf RQ1}: How does GOD perform compared with existing KD and self-supervised methods in SR? (Section~\ref{sec/overall_performance})
    \item {\bf RQ2}: Does GOD improve generalization and robustness under challenging conditions? (Section~\ref{sec/generalization_and_robustness})
    \item {\bf RQ3}: How effective are the core components of GOD? (Section~\ref{sec/ablation_study})
    \item {\bf RQ4}: How efficient and hyperparameter-sensitive is GOD in practice? (Section~\ref{sec/efficiency_and_sensitivity})
\end{itemize}

\subsection{Experimental Setup}

\begin{table}[t]
    \footnotesize
    \setlength{\tabcolsep}{3pt}
    \caption{Statistics of the datasets.}
    \label{tab/dataset}
    \begin{tabular}{c|ccccc}
        \toprule
        \textbf{Datasets} & \textbf{\# Interactions} & \textbf{\# Users} & \textbf{\# Items} & \textbf{Avg.SeqLen} & \textbf{Sparsity} \\
        \midrule
        Amazon Beauty & 198,502 & 22,363 & 12,101 &   8.9 & 99.93 \% \\
        Yelp          & 345,186 & 29,915 & 21,572 &  11.5 & 99.95 \% \\
        MovieLens 1M  & 999,611 &  6,040 &  3,416 & 165.5 & 95.16 \% \\
        \bottomrule
    \end{tabular}
\end{table}

\begin{table*}[t]
    \footnotesize
    \setlength{\tabcolsep}{2pt}
    \caption{Overall performance. Best and second-best results are marked in bold and underlined. $\boldsymbol{Improv.S}$ and $\boldsymbol{Improv.B}$ denote relative improvements over Student and the strongest baseline. Asterisk (*) indicates statistical significance at $p < 0.05$ by paired $t$-test over five independent runs against the strongest baseline.}
    \label{tab/performance}
    \begin{tabular}{c|c|c|c|ccccccc|ccc||c|cc}
        \toprule
        \textbf{Dataset} & \textbf{Backbone} & \textbf{Metric} & \textbf{Teacher} & \textbf{Student} & \textbf{RD} & \textbf{CD} & \textbf{DE} & \textbf{RRD} & \textbf{HTD} & \textbf{BD} & \textbf{AdaRec} & \textbf{MSKDIK} & \textbf{EMKD} & \textbf{GOD} & $\boldsymbol{Improv.S}$ & $\boldsymbol{Improv.B}$ \\
        \midrule\midrule
        \multirow{12}{*}{Amazon Beauty} & \multirow{4}{*}{GRU4Rec} & HR@10   & 0.0506 & 0.0384 & 0.0392 & 0.0403 & 0.0413 & 0.0426 & 0.0438 & 0.0464 & 0.0450 & 0.0454 & \underline{0.0472} & \textbf{0.0508*} & 32.29 \% &  7.63 \% \\
                                        &                          & HR@20   & 0.0788 & 0.0622 & 0.0626 & 0.0639 & 0.0652 & 0.0668 & 0.0682 & 0.0711 & 0.0692 & 0.0703 & \underline{0.0722} & \textbf{0.0777*} & 24.92 \% &  7.62 \% \\
                                        &                          & NDCG@10 & 0.0257 & 0.0187 & 0.0190 & 0.0196 & 0.0203 & 0.0210 & 0.0217 & 0.0232 & 0.0224 & 0.0227 & \underline{0.0237} & \textbf{0.0257*} & 37.69 \% &  8.44 \% \\
                                        &                          & NDCG@20 & 0.0322 & 0.0249 & 0.0262 & 0.0266 & 0.0270 & 0.0275 & 0.0281 & 0.0290 & 0.0285 & 0.0286 & \underline{0.0294} & \textbf{0.0316*} & 26.91 \% &  7.48 \% \\
        \cline{2-17}
                                        & \multirow{4}{*}{FMLPRec} & HR@10   & 0.0667 & 0.0524 & 0.0540 & 0.0547 & 0.0555 & 0.0564 & 0.0573 & 0.0583 & 0.0583 & 0.0590 & \underline{0.0596} & \textbf{0.0634*} & 20.99 \% &  6.38 \% \\
                                        &                          & HR@20   & 0.0959 & 0.0813 & 0.0843 & 0.0849 & 0.0856 & 0.0864 & 0.0871 & 0.0883 & 0.0873 & 0.0886 & \underline{0.0892} & \textbf{0.0951*} & 16.97 \% &  6.61 \% \\
                                        &                          & NDCG@10 & 0.0368 & 0.0296 & 0.0315 & 0.0319 & 0.0322 & 0.0326 & 0.0330 & 0.0336 & 0.0335 & 0.0338 & \underline{0.0341} & \textbf{0.0372*} & 25.68 \% &  9.09 \% \\
                                        &                          & NDCG@20 & 0.0436 & 0.0346 & 0.0363 & 0.0368 & 0.0372 & 0.0377 & 0.0382 & 0.0388 & 0.0388 & 0.0393 & \underline{0.0396} & \textbf{0.0427*} & 23.41 \% &  7.83 \% \\
        \cline{2-17}
                                        & \multirow{4}{*}{SASRec}  & HR@10   & 0.0727 & 0.0605 & 0.0617 & 0.0626 & 0.0635 & 0.0646 & 0.0655 & 0.0675 & 0.0666 & 0.0668 & \underline{0.0683} & \textbf{0.0725*} & 19.83 \% &  6.15 \% \\
                                        &                          & HR@20   & 0.1030 & 0.0925 & 0.0932 & 0.0943 & 0.0955 & 0.0969 & 0.0981 & 0.1006 & 0.0995 & 0.0999 & \underline{0.1017} & \textbf{0.1051*} & 13.62 \% &  3.34 \% \\
                                        &                          & NDCG@10 & 0.0401 & 0.0316 & 0.0321 & 0.0327 & 0.0332 & 0.0339 & 0.0346 & 0.0360 & 0.0353 & 0.0355 & \underline{0.0364} & \textbf{0.0392*} & 24.05 \% &  7.69 \% \\
                                        &                          & NDCG@20 & 0.0472 & 0.0390 & 0.0410 & 0.0415 & 0.0420 & 0.0426 & 0.0432 & 0.0444 & 0.0437 & 0.0439 & \underline{0.0447} & \textbf{0.0474*} & 21.54 \% &  6.04 \% \\
        \hline
        \multirow{12}{*}{Yelp}          & \multirow{4}{*}{GRU4Rec} & HR@10   & 0.0498 & 0.0365 & 0.0383 & 0.0379 & 0.0389 & 0.0391 & 0.0395 & 0.0403 & 0.0399 & 0.0404 & \underline{0.0407} & \textbf{0.0456*} & 24.93 \% & 12.04 \% \\
                                        &                          & HR@20   & 0.0843 & 0.0652 & 0.0686 & 0.0675 & 0.0701 & 0.0712 & 0.0725 & 0.0742 & 0.0739 & 0.0751 & \underline{0.0761} & \textbf{0.0810*} & 24.23 \% &  6.44 \% \\
                                        &                          & NDCG@10 & 0.0250 & 0.0187 & 0.0195 & 0.0191 & 0.0197 & 0.0201 & 0.0204 & 0.0208 & 0.0206 & 0.0210 & \underline{0.0212} & \textbf{0.0241*} & 28.88 \% & 13.68 \% \\
                                        &                          & NDCG@20 & 0.0337 & 0.0245 & 0.0263 & 0.0257 & 0.0266 & 0.0272 & 0.0277 & 0.0283 & 0.0278 & 0.0286 & \underline{0.0290} & \textbf{0.0322*} & 31.43 \% & 11.03 \% \\
        \cline{2-17}
                                        & \multirow{4}{*}{FMLPRec} & HR@10   & 0.0543 & 0.0407 & 0.0421 & 0.0429 & 0.0438 & 0.0443 & 0.0449 & 0.0460 & 0.0454 & 0.0456 & \underline{0.0464} & \textbf{0.0508*} & 24.82 \% &  9.48 \% \\
                                        &                          & HR@20   & 0.0873 & 0.0647 & 0.0676 & 0.0684 & 0.0700 & 0.0708 & 0.0715 & 0.0729 & 0.0723 & 0.0724 & \underline{0.0735} & \textbf{0.0804*} & 24.27 \% &  9.39 \% \\
                                        &                          & NDCG@10 & 0.0293 & 0.0223 & 0.0226 & 0.0231 & 0.0239 & 0.0244 & 0.0249 & 0.0258 & 0.0253 & 0.0255 & \underline{0.0262} & \textbf{0.0287*} & 28.70 \% &  9.54 \% \\
                                        &                          & NDCG@20 & 0.0375 & 0.0282 & 0.0296 & 0.0301 & 0.0311 & 0.0314 & 0.0316 & 0.0322 & 0.0320 & 0.0320 & \underline{0.0324} & \textbf{0.0360*} & 27.66 \% & 11.11 \% \\
        \cline{2-17}
                                        & \multirow{4}{*}{SASRec}  & HR@10   & 0.0568 & 0.0454 & 0.0481 & 0.0470 & 0.0485 & 0.0490 & 0.0494 & 0.0503 & 0.0497 & 0.0500 & \underline{0.0507} & \textbf{0.0567*} & 24.89 \% & 11.83 \% \\
                                        &                          & HR@20   & 0.0917 & 0.0764 & 0.0808 & 0.0791 & 0.0814 & 0.0818 & 0.0821 & 0.0828 & 0.0824 & 0.0826 & \underline{0.0831} & \textbf{0.0929*} & 21.60 \% & 11.79 \% \\
                                        &                          & NDCG@10 & 0.0303 & 0.0241 & 0.0259 & 0.0249 & 0.0263 & 0.0265 & 0.0267 & 0.0271 & 0.0269 & 0.0270 & \underline{0.0273} & \textbf{0.0311*} & 29.05 \% & 13.92 \% \\
                                        &                          & NDCG@20 & 0.0390 & 0.0316 & 0.0344 & 0.0334 & 0.0352 & 0.0356 & 0.0361 & 0.0370 & 0.0364 & 0.0367 & \underline{0.0373} & \textbf{0.0405*} & 28.16 \% &  8.58 \% \\
        \hline
        \multirow{12}{*}{MovieLens 1M}  & \multirow{4}{*}{GRU4Rec} & HR@10   & 0.2063 & 0.1325 & 0.1344 & 0.1334 & 0.1353 & 0.1361 & 0.1369 & 0.1375 & 0.1380 & 0.1386 & \underline{0.1391} & \textbf{0.1499*} & 13.13 \% &  7.76 \% \\
                                        &                          & HR@20   & 0.2976 & 0.1975 & 0.2030 & 0.1999 & 0.2046 & 0.2061 & 0.2076 & 0.2086 & 0.2096 & 0.2105 & \underline{0.2116} & \textbf{0.2231*} & 12.96 \% &  5.43 \% \\
                                        &                          & NDCG@10 & 0.1012 & 0.0665 & 0.0683 & 0.0671 & 0.0684 & 0.0686 & 0.0688 & 0.0689 & 0.0691 & 0.0690 & \underline{0.0692} & \textbf{0.0770*} & 15.79 \% & 11.27 \% \\
                                        &                          & NDCG@20 & 0.1260 & 0.0867 & 0.0895 & 0.0873 & 0.0909 & 0.0917 & 0.0924 & 0.0930 & 0.0934 & 0.0940 & \underline{0.0945} & \textbf{0.1029*} & 18.69 \% &  8.89 \% \\
        \cline{2-17}
                                        & \multirow{4}{*}{FMLPRec} & HR@10   & 0.1996 & 0.1338 & 0.1354 & 0.1349 & 0.1361 & 0.1363 & 0.1367 & 0.1369 & 0.1371 & 0.1373 & \underline{0.1375} & \textbf{0.1514*} & 13.15 \% & 10.11 \% \\
                                        &                          & HR@20   & 0.2894 & 0.2123 & 0.2167 & 0.2133 & 0.2200 & 0.2215 & 0.2231 & 0.2242 & 0.2253 & 0.2264 & \underline{0.2274} & \textbf{0.2398*} & 12.95 \% &  5.45 \% \\
                                        &                          & NDCG@10 & 0.0962 & 0.0694 & 0.0700 & 0.0696 & 0.0719 & 0.0729 & 0.0738 & 0.0745 & 0.0751 & 0.0757 & \underline{0.0764} & \textbf{0.0833*} & 20.03 \% &  9.03 \% \\
                                        &                          & NDCG@20 & 0.1205 & 0.0841 & 0.0884 & 0.0852 & 0.0890 & 0.0894 & 0.0897 & 0.0899 & 0.0901 & 0.0903 & \underline{0.0905} & \textbf{0.0992*} & 17.95 \% &  9.61 \% \\
        \cline{2-17}
                                        & \multirow{4}{*}{SASRec}  & HR@10   & 0.2330 & 0.1640 & 0.1664 & 0.1654 & 0.1683 & 0.1695 & 0.1707 & 0.1715 & 0.1723 & 0.1731 & \underline{0.1739} & \textbf{0.1856*} & 13.17 \% &  6.73 \% \\
                                        &                          & HR@20   & 0.3301 & 0.2490 & 0.2542 & 0.2513 & 0.2579 & 0.2598 & 0.2617 & 0.2630 & 0.2643 & 0.2655 & \underline{0.2668} & \textbf{0.2813*} & 12.97 \% &  5.43 \% \\
                                        &                          & NDCG@10 & 0.1150 & 0.0816 & 0.0839 & 0.0820 & 0.0856 & 0.0865 & 0.0873 & 0.0879 & 0.0885 & 0.0890 & \underline{0.0896} & \textbf{0.0978*} & 19.85 \% &  9.15 \% \\
                                        &                          & NDCG@20 & 0.1417 & 0.1052 & 0.1086 & 0.1058 & 0.1090 & 0.1091 & 0.1093 & 0.1094 & 0.1096 & 0.1097 & \underline{0.1098} & \textbf{0.1218*} & 15.78 \% & 10.93 \% \\
        \bottomrule
    \end{tabular}
\end{table*}

\subsubsection{Dataset}
To validate the effectiveness of GOD, we conduct experiments on three public datasets: Amazon Beauty \cite{mcauley2015image}, Yelp \cite{asghar2016yelp}, and MovieLens 1M \cite{harper2015movielens}.
These datasets vary in domain, scale, and sparsity, with detailed statistics provided in \autoref{tab/dataset}.
Following \cite{kang2018self, qiu2022contrastive}, we treat interactions as implicit feedback, and filter out users and items with fewer than five interactions.

\subsubsection{Baseline}
We first describe the backbone models to which GOD and all KD methods are applied.
We then summarize the KD baselines used for comparison.
\begin{itemize}[itemsep=1pt, left=0pt]
    \item \textbf{Backbone Recommender}:
        We apply GOD and all KD methods to GRU4Rec \cite{hidasi2015session}, FMLPRec \cite{zhou2022filter}, and SASRec \cite{kang2018self}, covering RNN-, MLP-, and Transformer-based architectures.
        For each backbone, Student is the compact model trained without KD, and Teacher is the larger pretrained model.
        GE is applied only to SASRec.
    \item \textbf{General Recommendation KD}:
        We compare with KD methods for general recommendation.
        RD \cite{tang2018ranking} transfers teacher ranking knowledge.
        CD \cite{lee2019collaborative} samples informative items from teacher rankings.
        DE \cite{kang2020rrd} distills latent embeddings.
        RRD \cite{kang2020rrd} relaxes teacher rankings for supervision.
        HTD \cite{kang2021topology} transfers hierarchical topology knowledge in the representation space.
        BD \cite{kweon2021bidirectional} performs bidirectional distillation for recommendation.
    \item \textbf{SR-tailored KD}:
        We further compare with KD designed for SR.
        AdaRec \cite{chen2021scene} searches for scene-adaptive student architectures.
        MSKDIK \cite{du2024multi} distills interest representation and drift knowledge in multiple stages.
        EMKD \cite{du2023ensemble} uses ensemble modeling with contrastive distillation.
\end{itemize}
We exclude LLM-based KD methods because they use external LLM knowledge or language-model students, while our experiments focus on ID-based KD with the same interaction-only input and backbone recommenders.\footnote{We discuss their relationship to GOD in Section~\ref{sec/knowledge_distillation}.}
This ensures that performance differences reflect the distillation strategy rather than additional modality or model-scale advantages.

\subsubsection{Evaluation}
For evaluation, we adopt the leave-one-out protocol \cite{kang2018self}, where the last item of each chronological sequence is used for testing, the penultimate item for validation, and the rest for training.
We perform full-ranking without negative sampling \cite{qiu2022contrastive}.
Performance is measured by HR@$k$ and NDCG@$k$, where $k \in \{10, 20\}$.

\subsubsection{Implementation}
We implement all models in PyTorch on a single NVIDIA GeForce RTX 3090 GPU and report averages over five independent runs.
Unless otherwise specified, the default student and teacher dimensions are $d^S = 16$ and $d^T = 64$, respectively.
For all backbones, we use 2 encoder layers, a dropout rate of 0.5, and a maximum sequence length $N = 50$.
For SASRec, we use 2 attention heads.
For fair comparison, we thoroughly tune each baseline following the search ranges in its original paper.
For GOD, the KD loss coefficient $\lambda$ is selected from \{1e-4, 1e-3, 1e-2, 1e-1, 1e-0\}, and the temperature $\tau$ from \{1, 5, 10, 20\}.
We train all models with Adam optimizer \cite{kingma2014adam} using a learning rate of 0.001 and batch size of 256.
Early stopping is applied with patience 30 based on validation NDCG@20.

\subsection{Overall Performance (RQ1)} \label{sec/overall_performance}

\autoref{tab/performance} reports the KD performance of GOD in SR.
Overall, GOD achieves the best results across all datasets, backbones, and metrics, improving over the strongest baseline by up to 13.92\%.
Beyond these gains, we make three key observations:
\begin{itemize}[itemsep=1pt, left=0pt]
    \item
        All KD methods outperform Student, confirming that teacher supervision helps compact models under sparse sequential feedback.
        SR-tailored KD methods generally outperform general recommendation KD methods, highlighting the importance of sequence-aware distillation.
    \item
        BD is a strong general KD baseline.
        On the sparser Amazon Beauty and Yelp datasets, BD often even surpasses AdaRec and MSKDIK, suggesting that adaptive teacher-student supervision benefits sparse and evolving interactions.
        However, unlike BD, GOD keeps the teacher frozen while still achieving better performance.
    \item
        GOD shows larger gains over Student on Amazon Beauty and Yelp than on the denser MovieLens 1M.
        This supports our motivation that grafting improves the generalization of distilled knowledge under sparse interactions, where SR models are prone to overfitting \cite{yao2021self, liu2021augmenting}.
        Consistent gains across GRU4Rec, FMLPRec, and SASRec further show the broad applicability of GOD across architectures.
\end{itemize}

\begin{table*}[t]
    \footnotesize
    \setlength{\tabcolsep}{3pt}
    \caption{Self-distillation performance with SASRec. Best and second-best results are marked in bold and underlined. $\boldsymbol{Improv.}$ denotes the relative improvement over the strongest baseline. Asterisk (*) indicates statistical significance at $p < 0.05$ by paired $t$-test over five independent runs against the strongest baseline.}
    \label{tab/self_distillation_performance}
    \begin{tabular}{c|c|c|ccc|cc|ccccc||c|c}
        \toprule
        \textbf{Dataset} & \textbf{Metric} & \textbf{SASRec} & \textbf{RRD} & \textbf{HTD} & \textbf{BD} & \textbf{MSKDIK} & \textbf{EMKD} & \textbf{CL4SRec} & \textbf{MCLRec} & \textbf{CT4Rec} & \textbf{DuoRec} & \textbf{RCL} & \textbf{GOD} & $Improv.$ \\
        \midrule\midrule
        \multirow{4}{*}{Amazon Beauty} & HR@10   & 0.0727 & 0.0718 & 0.0721 & 0.0749 & 0.0763 & 0.0785 & 0.0796 & 0.0806 & 0.0837 & \underline{0.0856} & 0.0853 & \textbf{0.0923*} & 7.83 \% \\
                                       & HR@20   & 0.1030 & 0.1010 & 0.1018 & 0.1047 & 0.1062 & 0.1078 & 0.1084 & 0.1099 & 0.1139 & 0.1164 & \underline{0.1172} & \textbf{0.1249*} & 6.57 \% \\
                                       & NDCG@10 & 0.0401 & 0.0396 & 0.0400 & 0.0426 & 0.0438 & 0.0454 & 0.0464 & 0.0474 & 0.0490 & 0.0506 & \underline{0.0517} & \textbf{0.0559*} & 8.12 \% \\
                                       & NDCG@20 & 0.0472 & 0.0461 & 0.0469 & 0.0499 & 0.0513 & 0.0525 & 0.0544 & 0.0554 & 0.0568 & 0.0580 & \underline{0.0596} & \textbf{0.0641*} & 7.55 \% \\
        \hline
        \multirow{4}{*}{Yelp}          & HR@10   & 0.0568 & 0.0559 & 0.0571 & 0.0574 & 0.0566 & 0.0584 & 0.0588 & 0.0591 & 0.0601 & 0.0606 & \underline{0.0607} & \textbf{0.0640*} & 5.44 \% \\
                                       & HR@20   & 0.0917 & 0.0903 & 0.0931 & 0.0934 & 0.0917 & 0.0945 & 0.0951 & 0.0957 & 0.0968 & 0.0975 & \underline{0.0978} & \textbf{0.1032*} & 5.52 \% \\
                                       & NDCG@10 & 0.0303 & 0.0296 & 0.0311 & 0.0311 & 0.0304 & 0.0304 & 0.0305 & 0.0306 & 0.0310 & \underline{0.0311} & 0.0310 & \textbf{0.0329*} & 5.79 \% \\
                                       & NDCG@20 & 0.0390 & 0.0382 & 0.0393 & 0.0397 & 0.0385 & 0.0391 & 0.0390 & 0.0396 & 0.0401 & \underline{0.0404} & 0.0403 & \textbf{0.0427*} & 5.69 \% \\
        \hline
        \multirow{4}{*}{MovieLens 1M}  & HR@10   & 0.2330 & 0.2266 & 0.2303 & 0.2318 & 0.2271 & 0.2336 & 0.2345 & 0.2304 & \underline{0.2352} & 0.2347 & 0.2308 & \textbf{0.2461*} & 4.63 \% \\
                                       & HR@20   & 0.3301 & 0.3236 & 0.3288 & 0.3285 & 0.3243 & 0.3317 & 0.3348 & 0.3275 & 0.3370 & \underline{0.3382} & 0.3342 & \textbf{0.3554*} & 5.09 \% \\
                                       & NDCG@10 & 0.1150 & 0.1133 & 0.1147 & 0.1155 & 0.1141 & 0.1177 & 0.1201 & 0.1188 & 0.1214 & 0.1230 & \underline{0.1232} & \textbf{0.1295*} & 5.11 \% \\
                                       & NDCG@20 & 0.1417 & 0.1390 & 0.1408 & 0.1419 & 0.1393 & 0.1442 & 0.1482 & 0.1455 & 0.1490 & 0.1497 & \underline{0.1500} & \textbf{0.1561*} & 4.07 \% \\
        \bottomrule
    \end{tabular}
\end{table*}

\paragraph{Self-Distillation}
\autoref{tab/self_distillation_performance} reports performance with SASRec when teacher and student have the same capacity ($d^S = d^T = 64$).
For GOD and KD baselines, this setting uses an offline pretrained teacher with the same architecture and dimension as the student, allowing us to examine whether GOD remains effective without a capacity gap.
Since no model compression is involved, we also compare with self-supervised contrastive baselines that use online signals generated by the model itself during training, such as augmented or consistency-based views: CL4SRec \cite{xie2022contrastive}, MCLRec \cite{qin2023meta}, CT4Rec \cite{chong2023ct4rec}, DuoRec \cite{qiu2022contrastive}, and RCL \cite{wang2024relative}.
GOD achieves the best results across all datasets and metrics, improving over the strongest baseline by up to 8.12\%.
We make three observations:
\begin{itemize}[itemsep=1pt, left=0pt]
    \item
        Conventional KD methods provide limited gains in this same-capacity setting.
        Several methods improve only marginally or even underperform SASRec.
        This suggests that, without a capacity gap, simply imitating a same-capacity teacher provides limited additional knowledge for recommendation.
    \item
        Self-supervised methods are stronger than KD baselines.
        This indicates that representation regularization is more effective than direct teacher imitation in the self-distillation setting.
        However, these methods rely on augmentation- or consistency-based views rather than teacher-conditioned source views.
    \item
        GOD consistently outperforms both KD and self-supervised baselines.
        This shows that its gains do not merely come from model compression or generic contrastive learning.
        Instead, grafted teacher-student views provide more transferable supervision to improve same-capacity sequence representation learning.
\end{itemize}

\subsection{Generalization and Robustness (RQ2)} \label{sec/generalization_and_robustness}

To further understand the generalization behavior of GOD, we evaluate its robustness under both data-side and teacher-side challenges.
We consider user history sparsity and interaction noise as data-side factors, and teacher-student capacity gap and teacher quality as teacher-side factors.

\begin{figure}[t]
  \centering
  \begin{subfigure}[t]{0.23\textwidth}
    \centering
    \includegraphics[width=\textwidth]{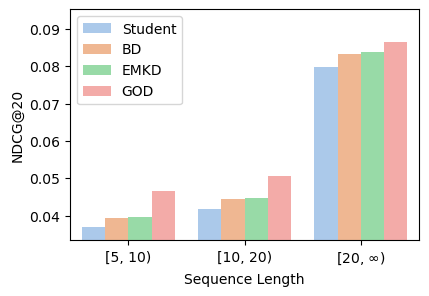}
    \caption{Amazon Beauty}
  \end{subfigure}
  \hfill
  \begin{subfigure}[t]{0.23\textwidth}
    \centering
    \includegraphics[width=\textwidth]{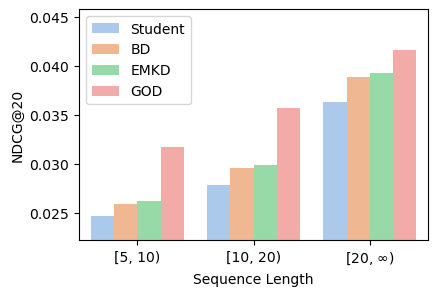}
    \caption{Yelp}
  \end{subfigure}
  \caption{Performance by sequence length with SASRec.}
  \label{fig/god_sparsity}
  \Description{}
\end{figure}

\paragraph{User History Sparsity}
We group users by sequence length and evaluate each group separately.
As shown in \autoref{fig/god_sparsity}, all methods improve with longer sequences, indicating that longer histories provide richer preference signals.
More importantly, GOD consistently performs best across all groups, with especially large margins in short-sequence groups.
This confirms that GOD is particularly effective when user interactions are sparse, where conventional KD methods have limited evidence for learning robust sequential patterns.
The margin becomes smaller for long-sequence users, suggesting that the benefit of grafting is most pronounced when generalization from sparse interactions is needed.

\begin{figure}[t]
  \centering
  \begin{subfigure}[t]{0.23\textwidth}
    \centering
    \includegraphics[width=\textwidth]{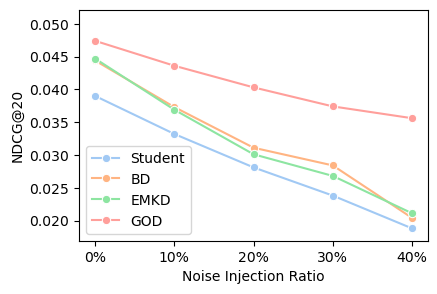}
    \caption{Amazon Beauty}
  \end{subfigure}
  \hfill
  \begin{subfigure}[t]{0.23\textwidth}
    \centering
    \includegraphics[width=\textwidth]{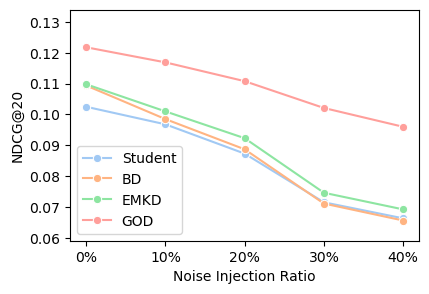}
    \caption{MovieLens 1M}
  \end{subfigure}
  \caption{Performance under different noise injection ratios with SASRec.}
  \label{fig/god_noise}
  \Description{}
\end{figure}

\paragraph{Noise Robustness}
We perturb the input sequence at test time by replacing a given ratio of items with random negative items, where the ratio is measured relative to the sequence length.
As shown in \autoref{fig/god_noise}, performance consistently drops as the noise ratio increases, confirming that noisy interactions make sequential pattern modeling more challenging.
Nevertheless, GOD maintains the best performance across all noise levels, and its performance gap remains clear even under high noise ratios.
This indicates that grafting provides more robust component-level supervision than bi-directional KD or SR-tailored KD, allowing the student to preserve transferable sequential patterns when observed histories are corrupted.

\begin{figure}[t]
  \centering
  \begin{subfigure}[t]{0.23\textwidth}
    \centering
    \includegraphics[width=\textwidth]{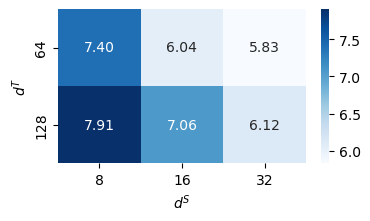}
    \caption{Amazon Beauty}
  \end{subfigure}
  \hfill
  \begin{subfigure}[t]{0.23\textwidth}
    \centering
    \includegraphics[width=\textwidth]{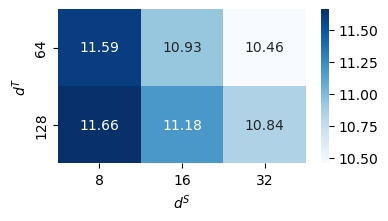}
    \caption{MovieLens 1M}
  \end{subfigure}
  \caption{Relative NDCG@20 improvements (\%) of GOD over EMKD, the strongest baseline, under different student and teacher dimensions with SASRec.}
  \label{fig/god_teacher_capacity}
  \Description{}
\end{figure}

\paragraph{Capacity Sensitivity}
We vary the student dimension $d^S \in \{ 8, 16, 32 \}$ and teacher dimension $d^T \in \{ 64, 128 \}$ to examine robustness of GOD across teacher-student capacity gaps.
As shown in \autoref{fig/god_teacher_capacity}, GOD consistently improves over EMKD across all dimension pairs.
The gains are larger when the student is smaller, indicating that grafting is especially useful when compact students lack sufficient capacity to absorb teacher knowledge.
In addition, increasing the teacher dimension generally brings larger improvements, suggesting that GOD can better exploit richer teacher knowledge instead of suffering from a wider teacher-student gap.
These results show that GOD remains effective across diverse compression ratios.

\begin{figure}[t]
  \centering
  \begin{subfigure}[t]{0.23\textwidth}
    \centering
    \includegraphics[width=\textwidth]{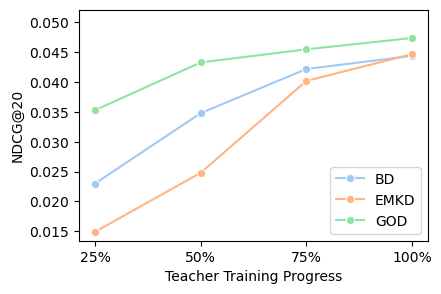}
    \caption{Amazon Beauty}
  \end{subfigure}
  \hfill
  \begin{subfigure}[t]{0.23\textwidth}
    \centering
    \includegraphics[width=\textwidth]{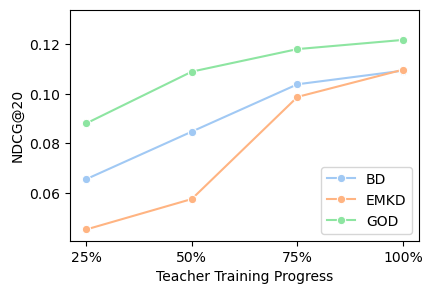}
    \caption{MovieLens 1M}
  \end{subfigure}
  \caption{Performance under different teacher training progress with SASRec. Here, 100\% denotes the best teacher checkpoint.}
  \label{fig/god_teacher_quality}
  \Description{}
\end{figure}

\paragraph{Teacher Quality Sensitivity}
We evaluate sensitivity to teacher quality by freezing checkpoints at 25\%, 50\%, 75\%, and 100\% of the training progress toward the best teacher checkpoint.
For example, if the best teacher checkpoint is obtained at epoch 44, the 50\% checkpoint is trained for 22 epochs.
As shown in \autoref{fig/god_teacher_quality}, all methods benefit from more trained teachers, confirming that teacher quality affects KD performance.
Notably, BD outperforms EMKD with weaker teachers, suggesting that bidirectional updates can partially compensate for weak teacher supervision.
GOD, however, consistently performs best without explicitly updating the teacher.
This indicates that grafting and GE provide adaptive teacher-conditioned supervision efficiently, enabling GOD to extract transferable knowledge even from partially trained teachers.

\subsection{Ablation Study (RQ3)} \label{sec/ablation_study}

To understand the contribution of each design in GOD, we conduct ablation studies on three core components: source models, GE, and GCL.

\begin{table}[t]
    \small
    \setlength{\tabcolsep}{3pt}
    \caption{Source model ablation with SASRec. NDCG@20 results are shown for the source models used in KD, where (T,T), (S,T), and (T,S) denote \textit{Non-Grafted Teacher}, \textit{Embed-Grafted Teacher}, and \textit{Encoder-Grafted Teacher}, respectively. Case (1) corresponds to GOD.}
    \label{tab/ablation_source_model}
    \begin{tabular}{cccc|ccc}
        \toprule
        \textbf{Case} & \textbf{(T,T)} & \textbf{(S,T)} & \textbf{(T,S)} & \textbf{Amazon Beauty} & \textbf{Yelp} & \textbf{MovieLens 1M} \\
        \midrule
        (1) & $\circ$ & $\circ$ & $\circ$ & \textbf{0.0474} & \textbf{0.0405} & \textbf{0.1218} \\
        \hline
        (2) & $\circ$ & $\circ$ &         &         0.0451  &         0.0375  &         0.1141  \\
        (3) & $\circ$ &         & $\circ$ &         0.0448  &         0.0376  &         0.1135  \\
        (4) &         & $\circ$ & $\circ$ &         0.0410  &         0.0342  &         0.1084  \\
        \hline
        (5) & $\circ$ &         &         &         0.0421  &         0.0349  &         0.1111  \\
        (6) &         & $\circ$ &         &         0.0401  &         0.0336  &         0.1074  \\
        (7) &         &         & $\circ$ &         0.0399  &         0.0337  &         0.1071  \\
        \bottomrule
    \end{tabular}
\end{table}

\paragraph{Source Models}
\autoref{tab/ablation_source_model} analyzes the contribution of each source model.
Case (1), corresponding to full GOD, achieves the best performance on all datasets.
Removing either hybrid source in cases (2) and (3) consistently degrades performance, confirming that \textit{Embed-Grafted Teacher} and \textit{Encoder-Grafted Teacher} provide complementary structural coupling.
The larger drop in case (4) shows that \textit{Non-Grafted Teacher} remains important as a stable knowledge anchor.
Cases (5), (6), and (7) further show that any single source alone is insufficient, indicating that GOD benefits from jointly distilling stable teacher knowledge and complementary grafted views.

\begin{figure}[t]
    \centering
    \includegraphics[width=0.7\linewidth]{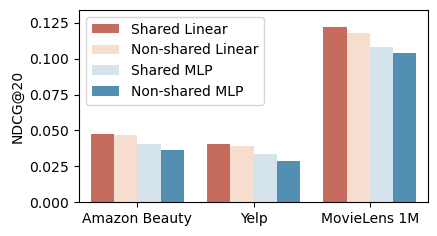}
    \caption{Projection design ablation with SASRec. Non-shared uses source model-specific projections, and MLP uses 2-layer projections. Shared Linear corresponds to GOD.}
    \label{fig/god_ablation_projection}
    \Description{}
\end{figure}

We further compare the shared linear projections (i.e., $W_{down}$ and $W_{up}$) used in GOD with non-shared linear projections and 2-layer MLP projections.
As shown in \autoref{fig/god_ablation_projection}, the shared linear design consistently achieves the best performance across all datasets.
Non-shared linear projections and MLP-based projections underperform despite their higher flexibility, indicating that the gains of GOD do not come from additional projection capacity.
This supports our design choice of using simple shared projections, which adapt dimensions while preserving the structural effect of grafting.

\begin{table}[t]
    \small
    \caption{GE ablation with SASRec. NDCG@20 is shown by varying teacher-student token interaction. T$\rightarrow$S only allows student-side query tokens to attend to teacher-side key/value tokens, while S$\rightarrow$T only allows the reverse direction. GE (half length) halves each side sequence length so that the concatenated sequence has the original length.}
    \label{tab/ablation_ge}
    \begin{tabular}{l|ccc}
        \toprule
        \textbf{Variant} & \textbf{Amazon Beauty} & \textbf{Yelp} & \textbf{MovieLens 1M} \\
        \midrule
        GE                   & \textbf{0.0474} & \textbf{0.0405} & \textbf{0.1218} \\
        T$\rightarrow$S only &         0.0466  &         0.0394  &         0.1184  \\
        S$\rightarrow$T only &         0.0459  &         0.0386  &         0.1166  \\
        GE (half length)     &         0.0462  &         0.0396  &         0.1152  \\
        w/o GE               &         0.0452  &         0.0379  &         0.1126  \\
        \bottomrule
    \end{tabular}
\end{table}

\paragraph{Grafted Encoding}
\autoref{tab/ablation_ge} evaluates GE variants, with full GE achieving the best performance on all datasets.
Uni-directional variants, i.e., T$\rightarrow$S only and S$\rightarrow$T only, outperform w/o GE, showing that cross-side token attention is useful even in one direction.
Among them, T$\rightarrow$S only performs better than S$\rightarrow$T only, suggesting that teacher-side tokens effectively stabilize student-side representations.
Notably, GE (half length) outperforms w/o GE, although its concatenated sequence length matches the original input length.
This shows that GE does not merely benefit from a longer attention span.
Finally, all partial variants remain inferior to full GE, indicating that bidirectional mutual attention with full-length histories is necessary to fully exploit teacher-student coupling.

\begin{table}[t]
    \small
    \caption{GCL ablation with SASRec. NDCG@20 is shown by varying pair construction and weighting. Source-Target only uses three pairs involving $h_u^{(S,S)}$, Unweighted uses $\mathcal{L}_{pair}$ instead of $\mathcal{L}_{kd}$, and Learnable weight replaces similarity-based weights with trainable pair weights.}
    \label{tab/ablation_gcl}
    \begin{tabular}{l|ccc}
        \toprule
        \textbf{Variant} & \textbf{Amazon Beauty} & \textbf{Yelp} & \textbf{MovieLens 1M} \\
        \midrule
        GCL                & \textbf{0.0474} & \textbf{0.0405} & \textbf{0.1218} \\
        Source-Target only &         0.0460  &         0.0386  &         0.1111  \\
        Unweighted         &         0.0455  &         0.0379  &         0.1090  \\
        Learnable weight   &         0.0427  &         0.0359  &         0.1074  \\
        \bottomrule
    \end{tabular}
\end{table}

\paragraph{Graft-aware Contrastive Learning}
\autoref{tab/ablation_gcl} analyzes the effect of GCL.
Full GCL achieves the best performance on all datasets, showing the benefit of adaptive pair-type weighting.
Source-Target only performs better than Unweighted, indicating that adding source-source pairs without proper weighting can introduce redundant or imbalanced supervision.
However, GCL further improves over Source-Target only, confirming that source-source pairs are useful when their contributions are dynamically controlled.
Learnable weight performs worst, suggesting that freely learnable pair weights are unstable and that similarity-based weighting with stop-gradient provides a more reliable distillation budget allocation.

\begin{figure}[t]
  \centering
  \begin{subfigure}[t]{0.23\textwidth}
    \centering
    \includegraphics[width=\textwidth]{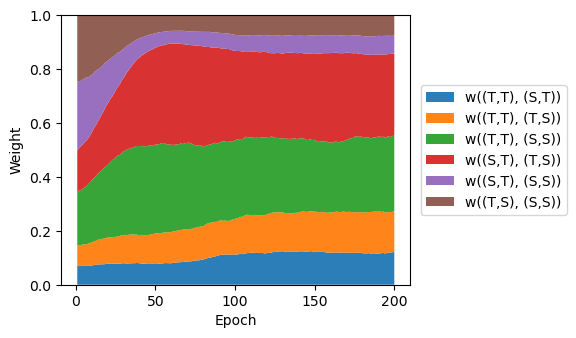}
    \caption{Amazon Beauty}
  \end{subfigure}
  \hfill
  \begin{subfigure}[t]{0.23\textwidth}
    \centering
    \includegraphics[width=\textwidth]{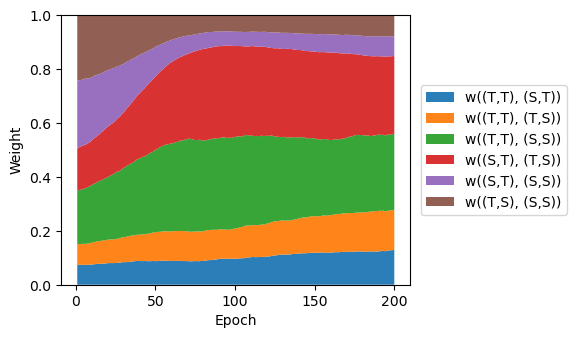}
    \caption{MovieLens 1M}
  \end{subfigure}
  \caption{Dynamics of six GCL pair weights over training. Higher pair similarity yields a lower weight.}
  \label{fig/god_gcl_weight}
  \Description{}
\end{figure}

To further understand GCL, we visualize the normalized weights of all six pairwise contrastive terms over training.
As shown in \autoref{fig/god_gcl_weight}, the weights do not collapse to a single pair but remain distributed across multiple relations, confirming that GCL performs balanced reweighting.
Early in training, pairs involving $h_u^{(S,S)}$ with grafted sources, i.e., $w((S,T), (S,S))$ and $w((T,S), (S,S))$, receive relatively high weights, while their importance gradually decreases as the student becomes better aligned with the sources.
In contrast, $w((T,T), (S,S))$ and $w((S,T), (T,S))$ gain higher weights and remain important in later epochs, indicating that GCL reallocates the distillation budget toward more informative and less redundant relations.

\subsection{Efficiency and Parameter Sensitivity (RQ4)} \label{sec/efficiency_and_sensitivity}

Beyond performance, practical KD requires efficient training and stable hyperparameter behavior.
We therefore analyze the training cost of GOD and its sensitivity to hyperparameters.

\begin{table}[t]
    \scriptsize
    \setlength{\tabcolsep}{2pt}
    \caption{Efficiency analysis with SASRec. Mem. denotes peak GPU memory, Time/Epoch denotes average training time per epoch, and Train denotes wall-clock training time until the best validation checkpoint.}
    \label{tab/efficiency}
    \begin{tabular}{c|cccc|cccc}
        \toprule
        \multirow{2}{*}{\textbf{Model}} & \multicolumn{4}{c|}{\textbf{Amazon Beauty}} & \multicolumn{4}{c}{\textbf{MovieLens 1M}} \\
        & Mem. & Time/Epoch & Train & NDCG@20 & Mem. & Time/Epoch & Train & NDCG@20 \\
        \midrule
        Teacher & 0.7 GB &  8.35 s &  6.12 m & 0.0472 & 1.1 GB & 15.88 s & 23.82 m & 0.1417 \\
        \hline
        Student & 0.6 GB &  7.75 s & 24.03 m & 0.0390 & 0.8 GB & 15.13 s & 18.66 m & 0.1052 \\
        BD      & 1.0 GB & 12.98 s & 27.69 m & 0.0444 & 1.5 GB & 26.67 s & 59.12 m & 0.1094 \\
        EMKD    & 1.1 GB & 15.57 s & 34.77 m & 0.0447 & 1.6 GB & 32.90 s & 51.00 m & 0.1098 \\
        \hline
        GOD     & 0.9 GB & 10.06 s & 15.26 m & 0.0474 & 1.2 GB & 21.00 s & 17.85 m & 0.1218 \\
        \bottomrule
    \end{tabular}
\end{table}

\paragraph{Efficiency}
\autoref{tab/efficiency} compares the training efficiency of GOD with representative baselines.
Although GOD incurs higher per-epoch time than Student due to hybrid source models and GE/GCL computation, it remains more efficient than BD and EMKD in both memory usage and time per epoch.
More importantly, by evaluating student embeddings and encoders through teacher-conditioned source models, GOD provides more direct and fine-grained supervision than output-level imitation.
This stronger supervision helps GOD reach its best validation performance in less wall-clock time than BD and EMKD, and even Student, despite the additional per-epoch computation.
At the same time, GOD achieves the best NDCG@20 among compact models.
These results show that GOD improves recommendation performance without prohibitive training overhead.

\begin{figure}[t]
    \centering
    \includegraphics[width=0.48\linewidth]{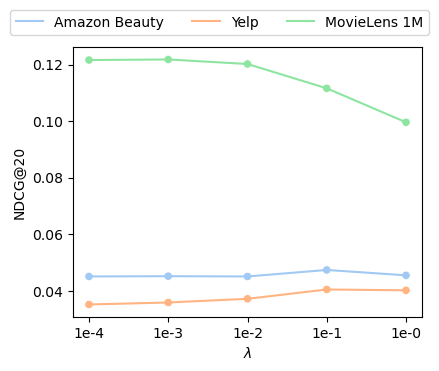}
    \includegraphics[width=0.48\linewidth]{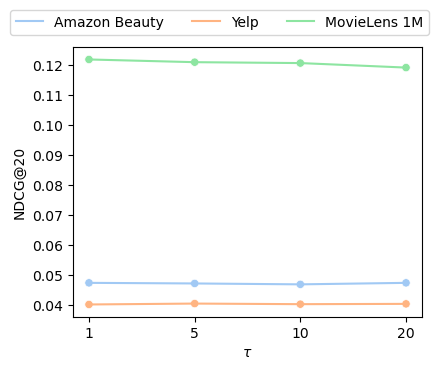}
    \caption{Parameter sensitivity with SASRec.}
    \label{fig/god_parameter}
    \Description{}
\end{figure}

\paragraph{Parameter Sensitivity}
We analyze the sensitivity of GOD to the loss coefficient $\lambda$ for $\mathcal{L}_{kd}$ and the temperature $\tau$ in $\mathcal{L}_{con}$.
As shown in \autoref{fig/god_parameter}, GOD is generally stable across a wide range of $\tau$, indicating that its contrastive distillation is not overly sensitive to temperature selection.
The effect of $\lambda$ is also moderate on Amazon Beauty and Yelp, while MovieLens 1M shows some degradation when $\lambda$ becomes too large.
This may suggest that excessive distillation strength can over-regularize the student on denser datasets \cite{kim2026tracer}, where recommendation supervision is already relatively sufficient.
Overall, GOD maintains robust performance under broad hyperparameter ranges, supporting its practical usability.

\section{Related Work}

\subsection{Knowledge Distillation} \label{sec/knowledge_distillation}

KD has been widely studied for recommendation to transfer knowledge from a high-capacity teacher to a compact student \cite{lee2021dual, chen2021scene, wei2024leave}.
Existing methods mainly differ in the form of supervision they distill.
Output-level approaches transfer teacher predictions or ranking preferences.
For example, RD \cite{tang2018ranking} distills teacher ranking lists, RRD \cite{kang2020rrd} relaxes ranking orders, and CD \cite{lee2019collaborative} selects informative items based on teacher rankings for collaborative supervision.
Representation-level approaches move beyond final rankings.
DE \cite{kang2020rrd} transfers latent embedding knowledge, while HTD \cite{kang2021topology} distills hierarchical topology in the representation space.
Beyond these fixed-teacher methods, BD \cite{kweon2021bidirectional} performs bidirectional distillation by updating the teacher-side knowledge source together with the student.
Together, these methods cover output-level, representation-level, and dynamic supervision, but still transfer teacher-produced signals after separate teacher-student execution rather than evaluating student components inside teacher-side computation.

In SR, KD must further transfer dynamic preference patterns from user interaction sequences, making sequence-aware supervision essential.
Recent SR-specific methods address this challenge from different perspectives \cite{du2024multi, xu2024slmrec, sun2025curriculum}.
AdaRec \cite{chen2021scene} searches for scene-adaptive student architectures and distills knowledge from sequential teachers.
MSKDIK \cite{du2024multi} distills interest representation and drift knowledge across multiple stages to capture evolving user intent.
EMKD \cite{du2023ensemble} leverages ensemble modeling with contrastive and logit-level distillation to provide richer sequential supervision.
These studies show that KD for SR benefits from modeling sequential dynamics beyond generic recommendation signals.
Nevertheless, they still distill teacher-produced outputs or representations as external targets, leaving student components outside teacher-side sequential computation.

Recently, another line of work leverages large language models (LLMs) for KD, aiming to transfer semantic understanding or reasoning ability to recommenders \cite{wang2024rdrec, wu2025bidirectional, zhang2025delrec}.
For example, DLLM2Rec \cite{cui2024distillation} distills LLM-derived content knowledge into ID-based sequential recommenders, while SLMRec \cite{xu2024slmrec} and SLIM \cite{wang2024can} distill LLM knowledge or reasoning processes for recommendation.
Although these approaches enrich recommendation with external semantic knowledge, they rely on massive LLM teachers, and some distilled students remain language models that are still much larger than ID-based sequential recommenders.
This leads to substantial distillation cost and can still incur higher inference latency than ID-based SR models.
In contrast, GOD focuses on KD between ID-based sequential recommenders.
Rather than injecting external semantic knowledge, GOD improves the generalization of distilled knowledge by providing component-level feedback through grafted teacher-student computation.

\subsection{Grafting}

Grafting was originally introduced as the inverse operation of pruning in decision trees \cite{webb1997decision} and later adapted to neural networks to replace or augment selected components.
For example, layer grafting \cite{meng2020filter} improves under-informative layers with validated external layers, NetGraft \cite{shen2021progressive} progressively replaces teacher layers with student counterparts for efficient few-shot KD, and GrafT \cite{park2024grafting} attaches add-on modules to vision transformers for multi-scale representation sharing.
These studies show that grafting can effectively modify selected components without redesigning the entire architecture.
Unlike prior grafting methods that mainly target model adaptation or component improvement, GOD uses grafting to construct teacher-conditioned source models for SR.
By replacing recommender-specific components, namely embedding tables and sequence encoders, GOD evaluates student components inside teacher-side sequential computation and provides component-level distillation feedback.

\section{Conclusion}

In this paper, we revisit KD for SR from the perspective of component-level feedback.
We identify a limitation of conventional distillation, where teacher-produced signals supervise the complete student path and can entangle the effects of student embeddings and encoders.
To address this, we propose GOD, a component-level KD framework that constructs teacher-conditioned source models through grafting.
By replacing selected frozen-teacher components with trainable student counterparts, GOD evaluates student components under teacher-side sequential computation and improves the generalization of distilled knowledge.
We further introduce GE and GCL to stabilize grafted representation generation and adaptively balance correlated source views.
Extensive experiments demonstrate that GOD consistently outperforms existing KD and self-supervised baselines, while maintaining practical training efficiency and using only the non-grafted student for inference.
Future work includes extending grafting to heterogeneous teacher-student architectures and exploring more flexible component-level knowledge transfer strategies.

\begin{acks}
    This work was supported by the NRF grant funded by the MSIT (No. RS-2024-00335873), the IITP grant funded by the MSIT (No.RS-2019-II191906, Artificial Intelligence Graduate School Program(POSTECH)).
\end{acks}

\appendix
\section{GenAI Usage Disclosure}

ChatGPT was used solely for English grammar and language refinement.
No generative AI was involved in the research process, including but not limited to coding, experiments, and data analysis.
All technical contributions are the original work of authors, and AI-assisted edits were manually reviewed to ensure alignment with the original intent.

\clearpage
\bibliographystyle{ACM-Reference-Format}
\balance
\bibliography{references}

@inproceedings{wang2015learning,
  title={Learning hierarchical representation model for nextbasket recommendation},
  author={Wang, Pengfei and Guo, Jiafeng and Lan, Yanyan and Xu, Jun and Wan, Shengxian and Cheng, Xueqi},
  booktitle={Proceedings of the 38th International ACM SIGIR conference on Research and Development in Information Retrieval},
  pages={403--412},
  year={2015}
}

@inproceedings{wu2019session,
  title={Session-based recommendation with graph neural networks},
  author={Wu, Shu and Tang, Yuyuan and Zhu, Yanqiao and Wang, Liang and Xie, Xing and Tan, Tieniu},
  booktitle={Proceedings of the AAAI conference on artificial intelligence},
  volume={33},
  number={01},
  pages={346--353},
  year={2019}
}

@inproceedings{zhou2020s3,
  title={S3-rec: Self-supervised learning for sequential recommendation with mutual information maximization},
  author={Zhou, Kun and Wang, Hui and Zhao, Wayne Xin and Zhu, Yutao and Wang, Sirui and Zhang, Fuzheng and Wang, Zhongyuan and Wen, Ji-Rong},
  booktitle={Proceedings of the 29th ACM international conference on information \& knowledge management},
  pages={1893--1902},
  year={2020}
}

@article{hidasi2015session,
  title={Session-based Recommendations with Recurrent Neural Networks},
  author={Hidasi, B},
  journal={arXiv preprint arXiv:1511.06939},
  year={2015}
}

@inproceedings{li2017neural,
  title={Neural attentive session-based recommendation},
  author={Li, Jing and Ren, Pengjie and Chen, Zhumin and Ren, Zhaochun and Lian, Tao and Ma, Jun},
  booktitle={Proceedings of the 2017 ACM on Conference on Information and Knowledge Management},
  pages={1419--1428},
  year={2017}
}

@inproceedings{tang2018personalized,
  title={Personalized top-n sequential recommendation via convolutional sequence embedding},
  author={Tang, Jiaxi and Wang, Ke},
  booktitle={Proceedings of the eleventh ACM international conference on web search and data mining},
  pages={565--573},
  year={2018}
}

@inproceedings{kang2018self,
  title={Self-attentive sequential recommendation},
  author={Kang, Wang-Cheng and McAuley, Julian},
  booktitle={2018 IEEE international conference on data mining (ICDM)},
  pages={197--206},
  year={2018},
  organization={IEEE}
}

@inproceedings{sun2019bert4rec,
  title={BERT4Rec: Sequential recommendation with bidirectional encoder representations from transformer},
  author={Sun, Fei and Liu, Jun and Wu, Jian and Pei, Changhua and Lin, Xiao and Ou, Wenwu and Jiang, Peng},
  booktitle={Proceedings of the 28th ACM international conference on information and knowledge management},
  pages={1441--1450},
  year={2019}
}

@inproceedings{xie2022contrastive,
  title={Contrastive learning for sequential recommendation},
  author={Xie, Xu and Sun, Fei and Liu, Zhaoyang and Wu, Shiwen and Gao, Jinyang and Zhang, Jiandong and Ding, Bolin and Cui, Bin},
  booktitle={2022 IEEE 38th international conference on data engineering (ICDE)},
  pages={1259--1273},
  year={2022},
  organization={IEEE}
}

@inproceedings{qiu2022contrastive,
  title={Contrastive learning for representation degeneration problem in sequential recommendation},
  author={Qiu, Ruihong and Huang, Zi and Yin, Hongzhi and Wang, Zijian},
  booktitle={Proceedings of the fifteenth ACM international conference on web search and data mining},
  pages={813--823},
  year={2022}
}

@inproceedings{tan2016improved,
  title={Improved recurrent neural networks for session-based recommendations},
  author={Tan, Yong Kiam and Xu, Xinxing and Liu, Yong},
  booktitle={Proceedings of the 1st workshop on deep learning for recommender systems},
  pages={17--22},
  year={2016}
}

@inproceedings{liu2018stamp,
  title={STAMP: short-term attention/memory priority model for session-based recommendation},
  author={Liu, Qiao and Zeng, Yifu and Mokhosi, Refuoe and Zhang, Haibin},
  booktitle={Proceedings of the 24th ACM SIGKDD international conference on knowledge discovery \& data mining},
  pages={1831--1839},
  year={2018}
}

@inproceedings{tuan20173d,
  title={3D convolutional networks for session-based recommendation with content features},
  author={Tuan, Trinh Xuan and Phuong, Tu Minh},
  booktitle={Proceedings of the eleventh ACM conference on recommender systems},
  pages={138--146},
  year={2017}
}

@inproceedings{ying2018sequential,
  title={Sequential recommender system based on hierarchical attention network},
  author={Ying, Haochao and Zhuang, Fuzhen and Zhang, Fuzheng and Liu, Yanchi and Xu, Guandong and Xie, Xing and Xiong, Hui and Wu, Jian},
  booktitle={IJCAI international joint conference on artificial intelligence},
  year={2018}
}

@inproceedings{li2020time,
  title={Time interval aware self-attention for sequential recommendation},
  author={Li, Jiacheng and Wang, Yujie and McAuley, Julian},
  booktitle={Proceedings of the 13th international conference on web search and data mining},
  pages={322--330},
  year={2020}
}

@inproceedings{fan2022sequential,
  title={Sequential recommendation via stochastic self-attention},
  author={Fan, Ziwei and Liu, Zhiwei and Wang, Yu and Wang, Alice and Nazari, Zahra and Zheng, Lei and Peng, Hao and Yu, Philip S},
  booktitle={Proceedings of the ACM web conference 2022},
  pages={2036--2047},
  year={2022}
}

@inproceedings{liu2023linrec,
  title={Linrec: Linear attention mechanism for long-term sequential recommender systems},
  author={Liu, Langming and Cai, Liu and Zhang, Chi and Zhao, Xiangyu and Gao, Jingtong and Wang, Wanyu and Lv, Yifu and Fan, Wenqi and Wang, Yiqi and He, Ming and others},
  booktitle={Proceedings of the 46th International ACM SIGIR Conference on Research and Development in Information Retrieval},
  pages={289--299},
  year={2023}
}

@inproceedings{yao2021self,
  title={Self-supervised learning for large-scale item recommendations},
  author={Yao, Tiansheng and Yi, Xinyang and Cheng, Derek Zhiyuan and Yu, Felix and Chen, Ting and Menon, Aditya and Hong, Lichan and Chi, Ed H and Tjoa, Steve and Kang, Jieqi and others},
  booktitle={Proceedings of the 30th ACM international conference on information \& knowledge management},
  pages={4321--4330},
  year={2021}
}

@inproceedings{liu2021augmenting,
  title={Augmenting sequential recommendation with pseudo-prior items via reversely pre-training transformer},
  author={Liu, Zhiwei and Fan, Ziwei and Wang, Yu and Yu, Philip S},
  booktitle={Proceedings of the 44th international ACM SIGIR conference on Research and development in information retrieval},
  pages={1608--1612},
  year={2021}
}

@inproceedings{wang2021denoising,
  title={Denoising implicit feedback for recommendation},
  author={Wang, Wenjie and Feng, Fuli and He, Xiangnan and Nie, Liqiang and Chua, Tat-Seng},
  booktitle={Proceedings of the 14th ACM international conference on web search and data mining},
  pages={373--381},
  year={2021}
}

@article{liu2021contrastive,
  title={Contrastive self-supervised sequential recommendation with robust augmentation},
  author={Liu, Zhiwei and Chen, Yongjun and Li, Jia and Yu, Philip S and McAuley, Julian and Xiong, Caiming},
  journal={arXiv preprint arXiv:2108.06479},
  year={2021}
}

@inproceedings{dang2023uniform,
  title={Uniform sequence better: Time interval aware data augmentation for sequential recommendation},
  author={Dang, Yizhou and Yang, Enneng and Guo, Guibing and Jiang, Linying and Wang, Xingwei and Xu, Xiaoxiao and Sun, Qinghui and Liu, Hong},
  booktitle={Proceedings of the AAAI conference on artificial intelligence},
  volume={37},
  number={4},
  pages={4225--4232},
  year={2023}
}

@inproceedings{qin2023meta,
  title={Meta-optimized contrastive learning for sequential recommendation},
  author={Qin, Xiuyuan and Yuan, Huanhuan and Zhao, Pengpeng and Fang, Junhua and Zhuang, Fuzhen and Liu, Guanfeng and Liu, Yanchi and Sheng, Victor},
  booktitle={Proceedings of the 46th International ACM SIGIR Conference on Research and Development in Information Retrieval},
  pages={89--98},
  year={2023}
}

@inproceedings{chong2023ct4rec,
  title={CT4Rec: Simple yet Effective Consistency Training for Sequential Recommendation},
  author={Chong, Liu and Liu, Xiaoyang and Zheng, Rongqin and Zhang, Lixin and Liang, Xiaobo and Li, Juntao and Wu, Lijun and Zhang, Min and Lin, Leyu},
  booktitle={Proceedings of the 29th ACM SIGKDD Conference on Knowledge Discovery and Data Mining},
  pages={3901--3913},
  year={2023}
}

@inproceedings{du2023ensemble,
  title={Ensemble modeling with contrastive knowledge distillation for sequential recommendation},
  author={Du, Hanwen and Yuan, Huanhuan and Zhao, Pengpeng and Zhuang, Fuzhen and Liu, Guanfeng and Zhao, Lei and Liu, Yanchi and Sheng, Victor S},
  booktitle={Proceedings of the 46th International ACM SIGIR Conference on Research and Development in Information Retrieval},
  pages={58--67},
  year={2023}
}

@inproceedings{wang2024relative,
  title={Relative Contrastive Learning for Sequential Recommendation with Similarity-based Positive Sample Selection},
  author={Wang, Zhikai and Shen, Yanyan and Zhang, Zexi and He, Li and Li, Yichun and Gu, Hao and Zhang, Yinghua},
  booktitle={Proceedings of the 33rd ACM International Conference on Information and Knowledge Management},
  pages={2493--2502},
  year={2024}
}

@article{hinton2015distilling,
  title={Distilling the Knowledge in a Neural Network},
  author={Hinton, Geoffrey},
  journal={arXiv preprint arXiv:1503.02531},
  year={2015}
}

@article{romero2014fitnets,
  title={Fitnets: Hints for thin deep nets},
  author={Romero, Adriana and Ballas, Nicolas and Kahou, Samira Ebrahimi and Chassang, Antoine and Gatta, Carlo and Bengio, Yoshua},
  journal={arXiv preprint arXiv:1412.6550},
  year={2014}
}

@inproceedings{tang2018ranking,
  title={Ranking distillation: Learning compact ranking models with high performance for recommender system},
  author={Tang, Jiaxi and Wang, Ke},
  booktitle={Proceedings of the 24th ACM SIGKDD international conference on knowledge discovery \& data mining},
  pages={2289--2298},
  year={2018}
}

@inproceedings{kang2020rrd,
  title={DE-RRD: A knowledge distillation framework for recommender system},
  author={Kang, SeongKu and Hwang, Junyoung and Kweon, Wonbin and Yu, Hwanjo},
  booktitle={Proceedings of the 29th ACM International Conference on Information \& Knowledge Management},
  pages={605--614},
  year={2020}
}

@inproceedings{kang2021topology,
  title={Topology distillation for recommender system},
  author={Kang, SeongKu and Hwang, Junyoung and Kweon, Wonbin and Yu, Hwanjo},
  booktitle={Proceedings of the 27th ACM SIGKDD Conference on Knowledge Discovery \& Data Mining},
  pages={829--839},
  year={2021}
}

@inproceedings{webb1997decision,
  title={Decision tree grafting},
  author={Webb, Geoffrey I},
  booktitle={IJCAI (2)},
  pages={846--851},
  year={1997}
}

@inproceedings{meng2020filter,
  title={Filter grafting for deep neural networks},
  author={Meng, Fanxu and Cheng, Hao and Li, Ke and Xu, Zhixin and Ji, Rongrong and Sun, Xing and Lu, Guangming},
  booktitle={Proceedings of the IEEE/CVF Conference on Computer Vision and Pattern Recognition},
  pages={6599--6607},
  year={2020}
}

@inproceedings{shen2021progressive,
  title={Progressive network grafting for few-shot knowledge distillation},
  author={Shen, Chengchao and Wang, Xinchao and Yin, Youtan and Song, Jie and Luo, Sihui and Song, Mingli},
  booktitle={Proceedings of the AAAI Conference on Artificial Intelligence},
  volume={35},
  number={3},
  pages={2541--2549},
  year={2021}
}

@inproceedings{kornblith2019similarity,
  title={Similarity of neural network representations revisited},
  author={Kornblith, Simon and Norouzi, Mohammad and Lee, Honglak and Hinton, Geoffrey},
  booktitle={International conference on machine learning},
  pages={3519--3529},
  year={2019},
  organization={PMLR}
}

@article{vaswani2017attention,
  title={Attention is all you need},
  author={Vaswani, A},
  journal={Advances in Neural Information Processing Systems},
  year={2017}
}

@article{devlin2018bert,
  title={Bert: Pre-training of deep bidirectional transformers for language understanding},
  author={Devlin, Jacob},
  journal={arXiv preprint arXiv:1810.04805},
  year={2018}
}

@article{oord2018representation,
  title={Representation learning with contrastive predictive coding},
  author={Oord, Aaron van den and Li, Yazhe and Vinyals, Oriol},
  journal={arXiv preprint arXiv:1807.03748},
  year={2018}
}

@inproceedings{lee2019collaborative,
  title={Collaborative distillation for top-N recommendation},
  author={Lee, Jae-woong and Choi, Minjin and Lee, Jongwuk and Shim, Hyunjung},
  booktitle={2019 IEEE International Conference on Data Mining (ICDM)},
  pages={369--378},
  year={2019},
  organization={IEEE}
}

@inproceedings{lee2021dual,
  title={Dual correction strategy for ranking distillation in top-n recommender system},
  author={Lee, Youngjune and Kim, Kee-Eung},
  booktitle={Proceedings of the 30th ACM International Conference on Information \& Knowledge Management},
  pages={3186--3190},
  year={2021}
}

@inproceedings{kweon2021bidirectional,
  title={Bidirectional distillation for top-K recommender system},
  author={Kweon, Wonbin and Kang, SeongKu and Yu, Hwanjo},
  booktitle={Proceedings of the Web Conference 2021},
  pages={3861--3871},
  year={2021}
}

@inproceedings{cui2024distillation,
  title={Distillation Matters: Empowering Sequential Recommenders to Match the Performance of Large Language Models},
  author={Cui, Yu and Liu, Feng and Wang, Pengbo and Wang, Bohao and Tang, Heng and Wan, Yi and Wang, Jun and Chen, Jiawei},
  booktitle={Proceedings of the 18th ACM Conference on Recommender Systems},
  pages={507--517},
  year={2024}
}

@article{kingma2014adam,
  title={Adam: A method for stochastic optimization},
  author={Kingma, Diederik P},
  journal={arXiv preprint arXiv:1412.6980},
  year={2014}
}

@inproceedings{saglietti2022solvable,
  title={Solvable model for inheriting the regularization through knowledge distillation},
  author={Saglietti, Luca and Zdeborov{\'a}, Lenka},
  booktitle={Mathematical and Scientific Machine Learning},
  pages={809--846},
  year={2022},
  organization={PMLR}
}

@inproceedings{zhang2018deep,
  title={Deep mutual learning},
  author={Zhang, Ying and Xiang, Tao and Hospedales, Timothy M and Lu, Huchuan},
  booktitle={Proceedings of the IEEE conference on computer vision and pattern recognition},
  pages={4320--4328},
  year={2018}
}

@inproceedings{zhou2022filter,
  title={Filter-enhanced MLP is all you need for sequential recommendation},
  author={Zhou, Kun and Yu, Hui and Zhao, Wayne Xin and Wen, Ji-Rong},
  booktitle={Proceedings of the ACM web conference 2022},
  pages={2388--2399},
  year={2022}
}

@inproceedings{chen2022intent,
  title={Intent contrastive learning for sequential recommendation},
  author={Chen, Yongjun and Liu, Zhiwei and Li, Jia and McAuley, Julian and Xiong, Caiming},
  booktitle={Proceedings of the ACM web conference 2022},
  pages={2172--2182},
  year={2022}
}

@inproceedings{qin2024intent,
  title={Intent contrastive learning with cross subsequences for sequential recommendation},
  author={Qin, Xiuyuan and Yuan, Huanhuan and Zhao, Pengpeng and Liu, Guanfeng and Zhuang, Fuzhen and Sheng, Victor S},
  booktitle={Proceedings of the 17th ACM international conference on web search and data mining},
  pages={548--556},
  year={2024}
}

@inproceedings{zhang2025frequency,
  title={Frequency-Augmented Mixture-of-Heterogeneous-Experts Framework for Sequential Recommendation},
  author={Zhang, Junjie and Xie, Ruobing and Lu, Hongyu and Sun, Wenqi and Zhao, Xin and Kang, Zhanhui and others},
  booktitle={THE WEB CONFERENCE 2025},
  year={2025}
}

@article{xu2024slmrec,
  title={SLMRec: Distilling large language models into small for sequential recommendation},
  author={Xu, Wujiang and Wu, Qitian and Liang, Zujie and Han, Jiaojiao and Ning, Xuying and Shi, Yunxiao and Lin, Wenfang and Zhang, Yongfeng},
  journal={arXiv preprint arXiv:2405.17890},
  year={2024}
}

@inproceedings{wang2024can,
  title={Can small language models be good reasoners for sequential recommendation?},
  author={Wang, Yuling and Tian, Changxin and Hu, Binbin and Yu, Yanhua and Liu, Ziqi and Zhang, Zhiqiang and Zhou, Jun and Pang, Liang and Wang, Xiao},
  booktitle={Proceedings of the ACM Web Conference 2024},
  pages={3876--3887},
  year={2024}
}

@inproceedings{park2024grafting,
  title={Grafting vision transformers},
  author={Park, Jongwoo and Kahatapitiya, Kumara and Kim, Donghyun and Sudalairaj, Shivchander and Fan, Quanfu and Ryoo, Michael S},
  booktitle={Proceedings of the IEEE/CVF Winter Conference on Applications of Computer Vision},
  pages={1145--1154},
  year={2024}
}

@inproceedings{zhou2023attention,
  title={Attention calibration for transformer-based sequential recommendation},
  author={Zhou, Peilin and Ye, Qichen and Xie, Yueqi and Gao, Jingqi and Wang, Shoujin and Kim, Jae Boum and You, Chenyu and Kim, Sunghun},
  booktitle={Proceedings of the 32nd ACM international conference on information and knowledge management},
  pages={3595--3605},
  year={2023}
}

@article{yuan2021improving,
  title={Improving sequential recommendation consistency with self-supervised imitation},
  author={Yuan, Xu and Chen, Hongshen and Song, Yonghao and Zhao, Xiaofang and Ding, Zhuoye and He, Zhen and Long, Bo},
  journal={arXiv preprint arXiv:2106.14031},
  year={2021}
}

@article{chen2021scene,
  title={Scene-adaptive knowledge distillation for sequential recommendation via differentiable architecture search},
  author={Chen, Lei and Yuan, Fajie and Yang, Jiaxi and Yang, Min and Li, Chengming},
  journal={arXiv preprint arXiv:2107.07173},
  year={2021}
}

@article{du2024multi,
  title={Multi-stage knowledge distillation for sequential recommendation with interest knowledge},
  author={Du, Yongping and Niu, Jinyu and Wang, Yuxin and Jin, Xingnan},
  journal={Information Sciences},
  volume={654},
  pages={119841},
  year={2024},
  publisher={Elsevier}
}

@inproceedings{wei2024leave,
  title={Leave no one behind: Online self-supervised self-distillation for sequential recommendation},
  author={Wei, Shaowei and Wu, Zhengwei and Li, Xin and Wu, Qintong and Zhang, Zhiqiang and Zhou, Jun and Gu, Lihong and Gu, Jinjie},
  booktitle={Proceedings of the ACM Web Conference 2024},
  pages={3767--3776},
  year={2024}
}

@inproceedings{mcauley2015image,
  title={Image-based recommendations on styles and substitutes},
  author={McAuley, Julian and Targett, Christopher and Shi, Qinfeng and Van Den Hengel, Anton},
  booktitle={Proceedings of the 38th international ACM SIGIR conference on research and development in information retrieval},
  pages={43--52},
  year={2015}
}

@article{asghar2016yelp,
  title={Yelp dataset challenge: Review rating prediction},
  author={Asghar, Nabiha},
  journal={arXiv preprint arXiv:1605.05362},
  year={2016}
}

@article{harper2015movielens,
  title={The movielens datasets: History and context},
  author={Harper, F Maxwell and Konstan, Joseph A},
  journal={Acm transactions on interactive intelligent systems (tiis)},
  volume={5},
  number={4},
  pages={1--19},
  year={2015},
  publisher={Acm New York, NY, USA}
}

@article{sun2025curriculum,
  title={Curriculum-scheduled knowledge distillation from multiple pre-trained teachers for multi-domain sequential recommendation},
  author={Sun, Wenqi and Xie, Ruobing and Zhang, Junjie and Zhao, Wayne Xin and Lin, Leyu and Wen, Ji-Rong},
  journal={World Wide Web},
  volume={28},
  number={6},
  pages={1--26},
  year={2025},
  publisher={Springer}
}

@inproceedings{wang2024rdrec,
  title={Rdrec: Rationale distillation for llm-based recommendation},
  author={Wang, Xinfeng and Cui, Jin and Suzuki, Yoshimi and Fukumoto, Fumiyo},
  booktitle={Proceedings of the 62nd Annual Meeting of the Association for Computational Linguistics (Volume 2: Short Papers)},
  pages={65--74},
  year={2024}
}

@article{wu2025bidirectional,
  title={Bidirectional Knowledge Distillation for Enhancing Sequential Recommendation with Large Language Models},
  author={Wu, Jiongran and Liu, Jiahao and Li, Dongsheng and Zhang, Guangping and Han, Mingzhe and Gu, Hansu and Zhang, Peng and Shang, Li and Lu, Tun and Gu, Ning},
  journal={arXiv preprint arXiv:2505.18120},
  year={2025}
}

@inproceedings{zhang2025delrec,
  title={DELRec: Distilling Sequential Pattern to Enhance LLMs-Based Sequential Recommendation},
  author={Zhang, Haoyi and Sun, Guohao and Lu, Jinhu and Liu, Guanfeng and Fang, Xiu Susie},
  booktitle={2025 IEEE 41st International Conference on Data Engineering (ICDE)},
  pages={1--14},
  year={2025},
  organization={IEEE}
}

@inproceedings{jiao2020tinybert,
  title={Tinybert: Distilling bert for natural language understanding},
  author={Jiao, Xiaoqi and Yin, Yichun and Shang, Lifeng and Jiang, Xin and Chen, Xiao and Li, Linlin and Wang, Fang and Liu, Qun},
  booktitle={Findings of the association for computational linguistics: EMNLP 2020},
  pages={4163--4174},
  year={2020}
}

@inproceedings{wang2019multi,
  title={Multi-similarity loss with general pair weighting for deep metric learning},
  author={Wang, Xun and Han, Xintong and Huang, Weilin and Dong, Dengke and Scott, Matthew R},
  booktitle={2019 IEEE/CVF Conference on Computer Vision and Pattern Recognition (CVPR)},
  pages={5017--5025},
  year={2019},
  organization={IEEE}
}

@inproceedings{sun2020circle,
  title={Circle loss: A unified perspective of pair similarity optimization},
  author={Sun, Yifan and Cheng, Changmao and Zhang, Yuhan and Zhang, Chi and Zheng, Liang and Wang, Zhongdao and Wei, Yichen},
  booktitle={2020 IEEE/CVF conference on computer vision and pattern recognition (CVPR)},
  pages={6397--6406},
  year={2020},
  organization={IEEE}
}

@inproceedings{kim2026flame,
  title={FLAME: Condensing Ensemble Diversity into a Single Network for Efficient Sequential Recommendation},
  author={Kim, WooJoo and Kim, JunYoung and Lim, JaeHyung and Choi, SeongJin and Kang, SeongKu and Yu, HwanJo},
  booktitle={Proceedings of the 49th International ACM SIGIR Conference on Research and Development in Information Retrieval},
  pages={823--833},
  year={2026}
}

@inproceedings{lee2026capturing,
  title={Capturing User Interests from Data Streams for Continual Sequential Recommendation},
  author={Lee, Gyuseok and Yoo, Hyunsik and Hwang, Junyoung and Kang, SeongKu and Yu, Hwanjo},
  booktitle={Proceedings of the Nineteenth ACM International Conference on Web Search and Data Mining},
  pages={313--323},
  year={2026}
}

@inproceedings{lim2025federated,
  title={Federated continual recommendation},
  author={Lim, Jaehyung and Kweon, Wonbin and Kim, Woojoo and Kim, Junyoung and Choi, Seongjin and Kim, Dongha and Yu, Hwanjo},
  booktitle={Proceedings of the 34th ACM International Conference on Information and Knowledge Management},
  pages={1798--1808},
  year={2025}
}

@article{kim2026overlooked,
  title={From Overlooked to Explored: Recovering Item Relations via Mixture of Perspectives for Sequential Recommendation},
  author={Kim, Junyoung and Kweon, Wonbin and Kim, Woojoo and Lim, Jaehyung and Kim, Dongha and Yu, Hwanjo},
  journal={arXiv preprint arXiv:2608.11846},
  year={2026}
}

@article{kim2026tracer,
  title={TRACER: Balancing Stability-Plasticity-Cognitivity Trilemma for LLM Enhanced Continual Recommendation},
  author={Kim, WooJoo and Yoo, HyunSik and Kim, JunYoung and Lim, JaeHyung and Kang, SeongKu and Yu, HwanJo},
  journal={arXiv preprint arXiv:2608.16075},
  year={2026}
}

\end{document}